\documentclass{pasj00}
\draft

\begin{document}
\SetRunningHead{Li Zhang et al.}{Multicolor Photometry of the
Galaxies in A 1775: Substructures, Luminosity Functions, and Star
Formation Properties}
\Received{}
\Accepted{}

\title{Multicolor Photometry of the Galaxies in Abell 1775:
Substructures, Luminosity Functions, and Star-Formation Properties}



%
 \author{
   Li \textsc{Zhang},\altaffilmark{1}
   Qirong \textsc{Yuan},\altaffilmark{1}
   Qian \textsc{Yang},\altaffilmark{2}
   Shiyan \textsc{Zhang},\altaffilmark{1}
   Feng \textsc{Li},\altaffilmark{3}
   Xu \textsc{Zhou},\altaffilmark{4}
   and
   Zhaoji \textsc{Jiang}\altaffilmark{4}
}
 \altaffiltext{1}{Department of Physics, Nanjing Normal University,
                 Wenyuan Road 1, Nanjing 210046, China}
 \email{lizhang722@163.com; yuanqirong@njnu.edu.cn}
 \altaffiltext{2}{Qingdao University, Qingdao 266071, China}
 \altaffiltext{3}{Jiangsu Industry College, Changzhou 213144, China}
 \altaffiltext{4}{National Astronomical Observatories, Chinese Academy of
   Sciences, Beijing 100012, China}


\maketitle

\begin{abstract}
An optical photometric observation in 15 bands was carried out for
nearby galaxy cluster Abell 1775 by the Beijing-Arizona-Taiwan-Connecticut
(BATC) multi-color system. Over 5000 sources' spectral energy
distributions (SEDs) were obtained. Since this cluster has also been
observed by the Sloan Digital Sky Survey (SDSS), the BATC SEDs were
combined with the SDSS five-band photometric data. Using the combined
SEDs, 146 faint galaxies were selected as new member galaxies by the
photometric redshift technique. Based on the positions, redshifts and
20-band SEDs of member galaxies, dynamical substructures and luminosity
functions (LFs) of A1775 were investigated. The previous reported bimodal
structure of A1775 has been confirmed: a poor subcluster with lower redshift,
A1775A, is located $\sim14'$ southeast to the main concentration A1775B.
After taking into account the new supplemented member galaxies, a new
subcluster A1775C was found along the aligned direction of A1775A and
A1775B. The different LF faint ends of the two subclusters indicate
that A1775B is a more dynamically evolved system, while A1775A is still
dynamically young. By the STARLIGHT spectral synthesis code, the star-
formation histories of the member galaxies were studied. The dependence
of the mean stellar ages upon the Hubble type was confirmed, and the
environmental effect on star-formation activities for galaxies in A1775B
has been explored.
\end{abstract}

\section{Introduction}
More and more galaxy clusters have been revealed to have
substructures by both the X-ray and optical surveys (Forman \& Jones
1982; Beers et al. 1991; Burns et al. 1994). Numerical simulation of
the evolution of galaxy clusters shows that at least $50\%$ of
apparently relaxed clusters contain significant substructures
(Salvador-Sol\'{e} et al. 1993). The dynamics of these ``lumpy''
clusters thus provide a means for exploring cluster evolution, which
may shed light on the theories of large-scale structure formation.
For example, the hierarchical scenario of structure formation believes
that massive clusters form through episodic merging of subunits, such
as galaxy groups and poor clusters, and through the continuous
accretion of field galaxies along the filaments (Zeldovich et al. 1982;
West et al. 1995; Kauffmann et al. 1999; Colberg et al. 2000). Galaxy
clusters also provide homogeneous samples of essentially coeval
galaxies in a high-density environment, which enable studies of
the evolution of stellar populations (Blakeslee et al. 2003; Poggianti
et al. 2008).

Abell 1775 (A1775; z=0.0717) is a richness 2 cluster of RS-type ``binary''
(Abell 1958; Struble \& Rood 1982; 1999). It has been observed in
optical, radio, and X-ray bands, showing a complicated picture of the
dynamics, substructure, and intergalactic medium. Its optical
observation shows that there are two giant elliptical galaxies at
the center of A1775 (Chincarini et al. 1971; Jenner 1974). These
two galaxies are considered to be a signature of merger events with
two less massive systems of galaxies. This inference is supported
by Oegerle, Hill, and Fitchett (1995, hereafter OHF) who presented
the velocity distribution of member galaxies in A1775. Although the
number of member galaxies in OHF is merely 51, their result clearly
indicates the presence of two subclusters along the line of sight.
Though no diffuse radio halo has been detected for this cluster, two
radio sources are found to be associated with the two dominant galaxies,
respectively. The south-east one, B1339+266B, is a head-tail radio
source, which exhibits a very narrow tail extended $\sim~8'$
northeast in the radio map (Giacintucci et al. 2007). In general,
the clusters containing narrow-angle tailed radio galaxies are
also believed to be dynamically complex systems undergoing merger
events (Bliton et al. 1998). The double peaks in the X-ray
brightness map of A1775 also show evidence for the cluster's
multiple nature (OHF; McMillan et al. 1989). The main intensive
X-ray peak is near the core emission of the tailed radio source,
which corresponds to the main concentration of A1775, and a small
clump of galaxies is also found around the other X-ray peak.
These intriguing observational features tell us that this cluster
is not a simple relaxed structure, but is still forming at the
present epoch.

This paper presents our new optical photometry for A1775 with the
Beijing-Arizona-Taiwan-Connecticut (hereafter BATC) multicolor
system. The BATC multicolor photometry can provide the information
about the spectral energy distributions (SEDs) for all of the objects
within a field of $58'\times58'$ centered on A1775. Accurate spectroscopic
and photometric data of A1775 have also been distributed by the
Sloan Digital Sky Survey (SDSS). For achieving a more accurate
estimate of photometric redshifts for detected galaxies, the SDSS
photometric data in five bands (Fukugita et al. 1996) are combined
with our BATC SEDs in fifteen bands. The availability of abundant
data facilitates the selection of faint member galaxies and enables
us to better understand the dynamical substructures, luminosity
functions, and the star-formation properties of the galaxy cluster A1775.

In section 2, the BATC multicolor photometry and data reduction are
presented, as well as a method for combining the SDSS SEDs with
the BATC SEDs. In section 3, using the galaxies with secure
spectroscopic redshifts in the field of view, the reliability of the
SED combination and the photometric redshift estimate is verified by
a comparison between the photometric redshift ($z_{ph}$) and the
spectroscopic redshift ($z_{sp}$). Further selection for faint
member galaxies based on their combined SEDs is also given in this
section. In section 4, the dynamics, spatial distribution, localized
velocity structure, luminosity function, and the star-formation
histories of the galaxies in A1775 are discussed. Finally, a summary
is made in section 5. Throughout this paper the {$\Lambda$CDM}
cosmology model is adopted with $H_0=70$kms$^{-1}$Mpc$^{-1}$,
$\Omega_m=0.3$, and $\Omega_\Lambda=0.7$.

\section{Observations and Data Reduction}
\subsection{Multicolor Photometry by the BATC System}
Observations of A1775 were carried out by the BATC multi-color
system with the 60/90 cm f/3 Schmidt telescope at the Xinglong
station, National Astronomical Observatories of China (NAOC). Before
2006 October, an old Ford CCD camera with a format of 2048$\times$2048
was used. The field of view was $58'\times58'$, with a scale of $1.''7$
pixel$^{-1}$. For pursuing a better spatial resolution and a higher
sensitivity in the blue bands, a new E2V 4096$\times$4096 thinned CCD
camera was equipped. The new CCD has a high quantum efficiency of
92.2\% at 4000\AA~and the field of view has been extended to $92'\times92'$
with a spatial scale of $1.''35$ pixel$^{-1}$. The pixel sizes for the
old and new CCDs are 15${\mu}m$ and 12${\mu}m$, respectively,
corresponding to a pixel size ratio of 5:4. 15 intermediate-band
filters covering the wavelength range from 3000 to 10000\AA~are
contained in the BATC filter system. These filters were especially
designed to avoid bright night sky emission lines (Fan et al. 1996).
The transmission curves of the BATC filters can be found in Figure 1
of Yuan Zhou, and Jiang (2003).

From 1996 to 2006, we accumulated 37 hr in only 12 bands, from d to p,
with the old CCD camera. In recent two years the exposures in the
a, b, c filters were completed with the new CCD camera. The observational
statistics is given in Table \ref{table1}. The total
exposure time reaches more than 48 hr. With an automatic
data-processing software, PIPELINE I (Fan et al. 1996), the standard
procedures of bias subtraction, flat-fielding correction, and position
calibration were carried out. The technique of integral pixel shifting
was used in the image combination, during which the cosmic rays and bad
pixels were removed by comparing multiple images.

For detecting and measuring the flux of sources within a given aperture
in the combined BATC images, a photometry package, PIPELINE II, developed
on the basis of DAOPHOT kernel (Stetson 1987; Zhou et al. 2003) was used.
An object was considered to be detected if its signal-to-noise ratio was
larger than 3.5 threshold in the i, j, and k bands. Considering that the
pixel size ratio between the old and new CCDs is 5:4, a radius of 4 pixels
for the images in 12 bands (from d to p), and a radius of 5 pixels for the
images in the other three bands (from a to c) were adopted as the
photometric apertures, respectively. A flux calibration in the h band was
performed using the Oke-Gunn primary flux standard stars HD 19445,
HD 84937, BD+26 2606, and BD+17 4708 (Oke \& Gunn 1983). To achieve
the relative SEDs of sources detected by the BATC system, Zhou et al. (1999)
developed a method of model calibration on the basis of the stellar SED
library. No calibration images of the standard stars are needed during the
flux calibration. Using this model calibration method, as a result, the SEDs
of more than 5000 sources have been obtained for further analysis.

For assessing the measurement errors at specified magnitudes, sources
are separated into different bins of magnitudes with an interval of 0.5
mag, and it is found that magnitude error in each filter is larger at
fainter depths. A typical error is less than 0.02 mag for those stars
brighter than 16.5 mag, and about 0.05 mag for stars with V $\sim18.5$
mag.

\subsection{Combining the SDSS and BATC SEDs}
Apart from the BATC photometric data, the SDSS photometric data were
also used in our analysis. For explicitly distinguishing the SDSS
filter names from those of the BATC, the SDSS filters and magnitudes
are referred to as $u'$, $g'$, $r'$, $i'$, and $z'$ in this paper,
which correspond to central wavelengths of 3560, 4680, 6180, 7500 and
8870\AA. From all of the SDSS-detected sources in our field of view,
the extended sources (galaxies) were extracted according to their
morphology classification given in the SDSS photometric catalog,
and were then cross-identified with objects detected in the BATC
observations. A searching circle with a radius of $2.'0$ centered at
the SDSS galaxies was adopted. By balancing the position offsets and
SED features of the counterparts, the identification is rather
unambiguous. Finally, 2248 galaxies with $r'<21.5$ mag detected by both
surveys were achieved.

Our previous studies show that the higher resolution of SED can improve
the accuracy of the photometric redshift estimate (Yuan et al. 2003;
Yang et al. 2004). For the SDSS photometric data of galaxies, following
the aperture correction method described in Yuan, Zhou, and Jiang (2003),
the SDSS colors of galaxies were derived by the same aperture of BATC,
that is an aperture of radius $r_{\rm ap}=6.''8$. For a given galaxy
detected by the SDSS imaging observation, the profile of its surface
brightness was quantified by various models. The total magnitude
estimated within 8 $r_{e}$ for deVaucouleurs profile or 4 $r_{e}$ for
exponential profile is termed the model magnitude, where $r_{e}$ is
the effective radius and its value can be found in the SDSS
photometric catalog. An aperture correction was applied to the SDSS
model magnitudes ($m_{\rm model}$) by
\begin{eqnarray}
\Delta m = m_{\rm ap} - m_{\rm model} = -2.5 \, log
\frac{\int_0^{r_{\rm ap}} 2 \pi r\, I(r) dr} {\int_0^{\infty}2\pi
r\, I(r) dr},
\label{equ1}
\end{eqnarray}
where $I(r)$ is the profile function of the surface intensity (e.g. the
exponential or deVaucouleurs $r^{1/4}$ model). The corresponding
parameters that quantify the preferred brightness profile for each
galaxy can be found in the SDSS photometric archive. The typical
values of $\Delta m$ estimated by equation (\ref{equ1}) is 0.0624.

For galaxies detected in both surveys, there might be an systematic
offset between the two photometric systems, which is called the zero
point. It can be determined by contrasting the BATC SEDs with the
aperture-corrected SDSS SEDs. The zero point is simply calculated
by averaging the magnitude differences at 6166\AA~and 7480\AA, the
effective wavelengths of the $r'$ and $i'$ filters. Interpolation
was performed during calculating the BATC magnitudes at 6166 and
7480\AA. This algorithm of zero point is slight different from that
in Liu et al. (2011). Figure \ref{zero-dis} presents the zero-point
distribution of 2248 galaxies with $r'<21.5$ mag detected by both
surveys. It can be seen that the SED zero-points are concentrated
at $\sim$ -0.2 mag. The zero points for different galaxies are slightly
different, since the deviation of the surface brightness profile from
the preferential model is different from source to source.

As a result, the combined SEDs of 2248 galaxies bright than $21.5$ mag
in the $r'$ band were obtained, which include the flux-calibrated BATC
SEDs in 15 filter bands and the aperture corrected SDSS SEDs in 5
filter bands. In the following analysis, we will focus on 881 galaxies
brighter than $19.5$ mag in the $r'$ band in the A1775 field.

\section{Selection of Faint Cluster Galaxies}
\subsection{Accuracy of Photometric Redshift}
Among the galaxies detected by both photometric surveys, 190
galaxies have been spectroscopically observed by the SDSS. This
spectroscopic sample provides us with an opportunity to test the
reliability of $z_{ph}$. For nearby galaxies, the most obvious and
useful spectral feature in redshift determination is the 4000\AA~
Balmer break, which ought to be better reflected by the combined
20-band SEDs. Especially when a galaxy has a large photometric
uncertainty in one filter of BATC, the magnitude in the nearby SDSS
band can be a good compensation, and vice-versa. Thus, the combined
photometric data of galaxies are used to get $z_{ph}$.

The SED fitting method ``Hyperz''(Bolzonella, Miralles, \& Pell\'{o}¡ä
2000) was performed for obtaining $z_{ph}$. Templates of normal
galaxies are taken into account. For correcting the intrinsic
extinction, the reddening law of the Milky Way (Allen 1976) was
adopted, and $A_{V}$ was allowed to vary in a range from 0.0 to
0.2, with steps of 0.02. The photometric redshift of a galaxy
was searched from 0.0 to 0.5, with an increment of 0.005. By
comparing the observed SED with the SEDs in the template library,
the best fit galaxy template was found, and its corresponding
redshift was taken as the $z_{ph}$ of the observed galaxy.
Meanwhile, the type of the observed galaxy was also ascertained.

Figure \ref{z-z} shows comparisons between $z_{ph}$ and $z_{sp}$ for
these 190 bright galaxies. $Z_{ph}$s in panel (a) were derived by 5
broad-band photometric data from SDSS, $z_{ph}$s in panel (b) were
derived by the combined SDSS and BATC SEDs. The solid lines in the
two panels are for $z_{ph}=z_{sp}$ and the dashed lines indicate
$z_{ph}=z_{sp}\pm0.05(1+z_{sp})$. The error bar of $z_{ph}$, which
corresponds to 68\% is also given. It can be seen that the $z_{ph}$
estimate is basically consistent with $z_{sp}$, and $z_{ph}$ in panel
(b), which was derived by combined photometric data, seems to have a
smaller dispersion. Many galaxies are concentrated at $z\sim0.07$.
A few catastrophic outliers also appear, it is likely that these
galaxies have peculiar features in internal dust extinction and
metallicity evolution. The corresponding distribution of offsets
$z_{ph}-z_{sp}$ for the above-two cases are given in figure \ref{zz-dis}.
A gaussian curve represented by a dashed line was used to fit the
distribution of $\Delta z$. The results showed that the centroid
of $\Delta z$ in panel (b) is closer to zero, and its dispersion
($\sigma_{\Delta z}$) is 0.007, which is smaller than that in
panel(a) ($\sigma_{\Delta z}=0.011$). This indicates that $z_{ph}$
we obtained is generally credible, and for most of the galaxies, their
$z_{ph}$s have been improved by combined multi-band photometric data.

\subsection{Cluster Membership of Faint Galaxies}
For a better understanding of the dynamical substructures in this cluster,
many faint member galaxies should be taken into account, but only a limited
number of bright member galaxies have been spectroscopically observed, for
instance, the limiting magnitude for the SDSS spectral survey is $r'=17.7$
mag, which corresponds to an absolute magnitude $M_{r'} \sim -20.0$ mag for
A1775 ($z=0.07$). To overcome this limit, the multicolor photometric data are
employed to complement faint member galaxies with $17.7$ mag $<r'<19.5$ mag.
After an exclusion of galaxies with secure spectroscopic redshifts, there are
657 galaxies brighter than $r'=19.5$ mag left. Based on the combined SEDs of
these galaxies, the photometric redshift technique is applied again with
the same galaxy templates and extinction law.

Figure \ref{zph-cand} gives the $z_{ph}$ distribution of these 657
galaxies. The peak at $z_{ph}\sim0.07$ corresponds to A1775. In order
to select faint galaxies belonging to A1775, the $z_{ph}$ value of
the former spectroscopic galaxies was applied for selecting likely
members. For spectroscopic galaxies with $0.05<z_{ph}<0.10$, the mean
value and the standard deviation of the $z_{ph}$ distribution are 0.073
and 0.008, respectively, which indicates a larger dispersion in the
$z_{ph}$ distribution with respect to the $z_{sp}$ distribution. In
order to decrease the contamination of foreground and background
galaxies, a strict 2$\sigma$ clipping algorithm (Yahil \& Vidal 1977)
was taken as the selection criterion. Faint galaxies with
$0.057(=0.073-2\times 0.008)<z_{ph}<0.089(=0.073+2\times0.008)$
have been picked out as possible member galaxies for further analysis.
As a result, 146 faint member candidates of A1775 were obtained. Based on
the sample of spectroscopic galaxies, the efficiency of the photometric
selection technique can be estimated. It is revealed that more than 85\%
of the members galaxies were selected by $z_{ph}$. The fraction of field
galaxy contamination due to errors in $z_{ph}$ is below 3\%. Table
\ref{table2} presents a catalog of the SED information of these newly
selected member galaxies, as well as their SDSS-given positions, $z_{ph}$
values, and morphology indices T (E, S0, Sa, Sb, Sc, Sd, and Im galaxies
are represented by 1 to 7, respectively).

\section{Properties of the Galaxies in A1775}
Within the BATC field of view, there are 224 galaxies with known
spectroscopic redshifts; among them, 190 galaxies with $14.5$ mag
$<r'<17.7$ mag have been spectroscopically observed by the SDSS, and the
remaining 34 bright galaxies are listed in the NASA/IPAC
Extragalactic Database (NED) only. Figure \ref{cz}a shows the $z_{sp}$
distribution of these 224 galaxies. 151 galaxies with $0.06< z_{sp}<
0.085$ were selected as probable member galaxies of A1775. After
applying the 2$\sigma$ clipping algorithm (Yahil \& Vidal 1977), no
galaxies were excluded. Thus, it is unambiguous to regard these 151
galaxies as spectroscopically confirmed member galaxies. These
member galaxies are referred to as ``sample I'' (see table \ref{table3}).
Combining with 146 faint member candidates newly selected, an enlarged
sample of 297(=151+146) member galaxies is constructed to study the
substructures and luminosity functions, to which we refer to
as ``sample II''.

\subsection{The KMM Partition and Velocity Distribution}
The profile of the line-of-sight velocity distribution is a useful
tool for investigating the dynamics of galaxy clusters (Quintana et
al. 1996; Muriel et al. 2002). The possible deviations from a Gaussian
distribution of the cluster galaxies might provide important
indications of substructure and ongoing merger. Figure \ref{cz}b
shows the corresponding velocity distribution of the galaxies in
Figure \ref{cz}a. The bimodal velocity distribution clearly implies that
there are two subclusters in A1775. Following OHF, the low-redshift
subcluster is referred to as A1775A, and the subcluster with high
redshift is referred to as A1775B. The velocities of the two central
giant elliptical galaxies are marked with arrows. Although the two
galaxies appears to be close to each other in optical images, their
radial velocities are significantly different.

To separate member galaxies of A1775A and A1775B more reliably ,
a prevalent partition method, KMM algorithm, was applied to the
151 member galaxies in sample I on the basis of their positional
and redshift information. The KMM is a maximum-likelihood algorithm
which assigns objects into groups and assesses the improvement in
fitting a multi-group model over a single group model (Ashman et al.
1994; Nemec \&Nemec 1993). After setting the initial parameters
of each subcluster, including estimated mean values and standard
deviations of the position and radial velocity, the solution of the
KMM algorithm was obtained: there are 49 galaxies belonging to A1775A,
and 102 galaxies belonging to A1775B. It was
noticed that the two central dominant galaxies had been divided into
different subclusters. The galaxy, SDSS J134149.14+2622224.5, with
a radial velocity of 22704 km$s^{-1}$, is similar to the mean
velocity of A1775B, and is thus assigned to A1775B. The other giant
elliptical galaxy, SDSS J134150.45+262213.0, is assigned to A1775A
with a membership possibility of 65.5\%. Its radial velocity is
20812 km$s^{-1}$, which significantly deviates from the velocity
centroid of A1775A.

Figure \ref{rest-cz} give the rest-frame velocity distribution of
member galaxies in A1775A and A1775B. To characterize the kinematical
properties of these two subclusters, two robust estimators, namely
the biweight location ($C_{BI}$) and scale ($S_{BI}$), was used, which
are defined by Beers Flynn, and Gebhardt (1990). These two quantities
are analogous to the mean value and standard deviation, and they are
robust for a broad range of probable non-Gaussian underlying populations
because of their insensitivity to outliers. For 49 galaxies in
A1775A, we obtained $C_{BI}=19571\pm67km^{-1}$ and $S_{BI}=432\pm
64kms^{-1}$; for 102 galaxies in A1775B, we got $C_{BI}=22551\pm
66km^{-1}$ and $S_{BI}=614\pm65km^{-1}$. The velocity difference
between the two subclusters is about $2780\pm88kms^{-1}$ in the
rest-frame of A1775. The previous OHF's results for A1775A are
$\overline{v}=19462kms^{-1}$, $\sigma=518kms^{-1}$; for A1775B they are
$\overline{v}=22767kms^{-1}$, $\sigma=394kms^{-1}$. Compared with
their results, the centroid velocities of the two subclusters that we
obtained are basically consistent with theirs, but the intrinsic
dispersion of A1775B that we gained is larger. Since the spectroscopically
confirmed member galaxies that we used were twice more than that which
they used, we believe our results are more reliable.

\subsection{Spatial Distribution}
Figure \ref{surfdence}a shows the spatial distribution of the 151
galaxies in sample I with respect to the central position of A1775
(13h41m55.6s, +26d21m53s; J2000.0, NED-given), superposed with the
contour map of the surface density which was smoothed by a Gaussian
window with $\sigma=2'$. Two subclusters are clearly
presented: the main concentration corresponds to the high-velocity
subcluster A1775B (member galaxies of it are denoted by ``$\bullet$''),
and the low-velocity subcluster A1775A (its member galaxies are
denoted by ``$\bigtriangleup$'') is located at about $14'$
southeast of A1775B. Two peaks of X-ray emission detected by ROSAT
are marked with ``$\oplus$''. Two central giant elliptical galaxies
taken as a galaxy pair by Chincarini et al.(1971) are marked with
``$+$''. A rich cluster with double dominant galaxies preferentially
indicates an ongoing merger event in this cluster. The well-studied
nearby rich cluster, Coma, is a good example (Colless \& Dunn 1996).

It is interesting that a narrow-angle tail radio source is found
to be exactly associated with the southeast dominant galaxy whose
velocity significantly deviates from A1775A, but is assigned to
A1775A (Giacintucci et al. 2007). Generally speaking, a head-tail
radio galaxy is regard as being the most striking example of the
interaction between the intra-cluster medium (ICM) and radio sources.
According to the viewpoint of Bliton et al.(1998), those clusters
containing narrow-angle tailed radio galaxies are dynamically complex
systems undergoing merger events. The soft X-ray band observation of
A1775 by ROSAT realed that there are two X-ray peaks. The main X-ray
peak is basically associated with the center of subcluster A1775B,
while the fainter X-ray peak has some offset from the galaxy density
peak of A1775A and it is offset towards the gravitational potential
of A1775B. This might also be a signature of subcluster interacting.
It may indicate that the clump A1775A is in the process of falling
into A1775B.

Figure \ref{surfdence}b shows the spatial distribution of the
galaxies in sample II, including the newly selected faint member
galaxies (denoted by ``$\circ$''). After supplementing, 146 faint
galaxies, the clump A1775A became more significant; and two new
clumps were found: one is located at $\sim15'$ NW of the main
concentration, namely A1775C, just along the direction aligned
with A1775A and A1775B; the other, namely A1775D, is at $\sim16'$
SW of A1775B. Besides, the spatial distribution of the galaxies in
the inner region of A1775B, defined by a surface density larger than
0.35 arcmin$^{-2}$, seems to be in alignment with A1775A, A1775B,
and A1775C, which may reflect some helpful clues about the dynamical
evolution history of A1775. However, because no abnormities around
the positions of A1775C and A1775D have been found in both the X-ray
and radio bands, the presence of A1775C and A1775D can't eliminate
the projection effect and it needs to be verified by follow-up
spectroscopy of faint member galaxies. In our following dynamical
analysis, the $\kappa-test$ algorithm will be used to check if the
new discovered structures A1775C and A1775D are real galaxy clumps.

On the whole, the bimodal structure in A1775 along the NW-SE
direction has been unveiled in both optical and X-ray bands.
Coincidentally, the connection direction of the two subclusters
(A1775A and A1775B) is just consistent with the aligned direction
of the central double galaxy, which may directly give us some clues
about the evolution of this cluster; the subclusters A1775A and A1775B
might have been interacted, and the original dominant galaxy in
A1775A seems to be ``pulled out'' by the gravitation of the main
subcluster, A1775B. This made the velocity of the galaxy significantly
deviate from the centroid velocity of A1775A, and enabled the galaxy
to interact with the ICM. It thus formed a head-tail radio source.

\subsection{Localized Velocity Structure}
To probe the robustness of the above subcluster detections, the
$\kappa-test$ (Colless \& Dunn 1996) was performed to quantify the
localized variation in the velocity distribution for both sample I
and sample II. The $\kappa-test$ is sensitive to spatially compact
subsystems that have either an average velocity that differs from
the cluster mean, or a velocity dispersion that differs from the
global one, or both. Keeping the spirit of the Dressler-Shectman
test (Dressler\& Shectman 1988), Colless \& Dunn (1996) defined
a statistic, $\kappa_{n}$, as follows:
$\kappa_n=\sum\limits_{i=1}^{N}{-log[P_{KS}(D>D_{obs})]}$, where n
is the size of the local group (n nearest neighbor galaxies), $N$
is the number of all member galaxies of a cluster, and $D$ is the
statistic in the standard K-S test. The statistic $\kappa_{n}$
expresses the accumulated difference between the local velocity
distribution and the whole cluster velocity distribution. The
probability of the substructure detection can be estimated by a
large number of Monte Carlo simulations.

For sample I, the presence of two subclusters in A1775 is strongly
supported. For each size of the local group, $10^{3}$ Monte Carlo
simulations were performed to estimate the probability of substructures
under this situation. As shown in table \ref{table4}, in a wide
range of local sizes, the value of $\kappa_{n}>\kappa_{n}^{obs}$ is
lower than 1\%, which means that the probability of the substructures
existence is more than 99\%. Among them, the optimum neighbor size
is n=9, for which all of the simulated cases are found to have
$\kappa_{9}$ values smaller than the observed case ($\kappa_{9}^{obs}$).
That is to say the probability of substructures existence in this case
is nearly 100\%. Figure \ref{bubble}a shows the local velocity
deviation for galaxies in sample I with a bubble plot. The
bubble size for each galaxy is proportional to $-log[P_{KS}(D>D_{obs})]$,
which reflects the probability of the local velocity distribution
different from the whole cluster velocity distribution in this place.
The larger is the bubble, the greater is the difference between the local
and overall velocity distributions. A comparison with Figure \ref{surfdence}
a shows that the bubble clustering indeed appears at the positions
of clumps A1775A and A1775B, and the probability of substructure
detection is nearly 100\%.

When we applied the $\kappa-test$ on sample II, the probability of
substructure detection was still great, with $P(\kappa_7 >
\kappa_7^{obs}) = 0.1\%$. The degree of the difference between the
local and overall velocity distributions for groups of 7 nearest
neighbors is shown by a bubble plot in Figure \ref{bubble}b.
Compared with Figure \ref{bubble}a, the bubble sizes in A1775A
and A1775B appear to be smaller, which is likely due to the photometric
redshift errors of the newly selected members. The double structures
detected with the spectroscopically confirmed member galaxies have
been, to some extent, smoothed in the velocity domain. Besides, a bunch
of bubbles is present at the position of the clump A1775C; however,
no bubble clustering can be found at the A1775D position. This
implies that A1775C might be a real clump, while A1775D might
be an unreal substructure which is due to the projection effect.

\subsection{Luminosity Functions of Subclusters}
The luminosity function (LF) in a cluster is a key observational
diagnostic for studying the formation and evolution of galaxies
in a dense environment. Assuming that galaxy mass-to-light ratios are
nearly constant for similar types of galaxies, the LF can potentially
provide a direct link to the initial mass function and hence the
distribution of the density perturbations that are thought to give rise
to galaxies (Press \& Schechter 1974). Previous studies showed that
the BM type and the core or outskirts region of the cluster may all
affect the galaxy LF (Biviano et al. 1995; Durret, Adami \& Lobo 2002;
Yang et al. 2004). The LFs of galaxy clusters have been well described
by the Schechter function, which was first proposed by Schechter (1976):
\begin{equation}
\phi(M)dM=\phi^*10^{0.4(\alpha+1)(M^*-M)}exp[-10^{0.4(M^*-M)}]dM,
\end{equation}
where $\phi^*$, $M^*$, and $\alpha$ are the normalization parameter,
the characteristic absolute magnitude, and the slope parameter at
faint tail, respectively.

Although the LF of cluster A1775 as a whole was studied by
Barkhouse et al. (2007), we are the first to explore the LFs
of the subclusters A1775A and A1775B separately. For the
spectroscopically confirmed member galaxies in sample I, the SDSS
spectroscopy is down to $r'=17.77$ mag, corresponding to a limit
of $M_{r'}\sim -20.0$ mag for A1775. These bright member galaxies
alone are considered to be insufficient to constrain the LF shape.
Fortunately, the benefit from the supplemented member galaxies
selected by $z_{ph}$, is that the faint parts of the LFs can be
better constrained.
These galaxies are selected by a criterion of being bright than
$r'=19.5$ mag, which corresponds to an absolute magnitude limit of
$M_{r'}\sim -18.0$ mag for A1775. It is about 2.0 mag deeper than the
spectroscopic galaxy sample. To reveal the luminosity function of
each subcluster, the newly-selected member galaxies were separated
into two subclusters. Considering the poor precision of the $z_{ph}$
estimate, we elected to perform the KMM partition algorithm on the
basis of only the position information. As a result, 55 galaxies
are assigned to A1775A, and 91 galaxies are assigned to A1775B.
The overall rate of correct allocation is estimated to be over 94\%.

Figure \ref{lf-AB} shows the LFs for galaxies in the
two subclusters, with an $M_{r'}$ bin of 0.5 mag. The dashed line
means SDSS spectroscopy completeness. For the poorer subcluster,
A1775A (panel a), the best fit with a single Schechter function is
$M^*_{r'}=-22.06\pm0.30$ mag, and $\alpha=-1.00\pm0.08$. A flat
galaxy count is extended to the faint end at $M_{r'}=-18.0$ mag.
However, for the main concentration A1775B (panel b), the best fit
of LF with Schechter function shows $M^*_{r'}=-20.75\pm0.10$ mag,
and $\alpha=-0.38\pm0.08$, which exhibits a deficit of galaxies
fainter than $M_{r'}=-19.0$ mag. Meanwhile, there are many relatively
bright galaxies between $M_{r'}=-21.5$ mag and $-19.5$ mag in this
subcluster. The difference of LFs between the two subclusters is
rather significant, which is due to the different galaxy contents
of these two regions. A1775B has a higher fraction of bright/massive
galaxies, and a drop can be expected near $M_{r'}>-18.0$ mag. For the
poor and loose subcluster A1775A, faint galaxies are predominant.
The observed paucity of faint galaxies in A1775B can be partially
explained by the fact that faint galaxies may have been swallowed
by relatively bright galaxies. Hence the richer system A1775B might
be a more dynamically evolved system and the poorer systems A1775A
might evolve marginally slower.

Since BATC photometry is complete down to $h=20.0$ mag (see Table
\ref{table1}), all member galaxies brighter than $r'=19.5$ mag
corresponding to $M_{r'}\sim -18.0$ mag should have been detected by
BATC. However, the cluster members we added in analyzing the LFs
were selected by the $2\sigma$ clipping method and since there is a
bias between $z_{sp}$ and $z_{ph}$, the completeness of members galaxies
in sample II down to $M_{r'}\sim -18.0$ mag may not reach 100\%.
Conservatively, at least 85\% of the member galaxies are included in this
galaxy sample and the estimated fraction of field galaxy contamination
is below 3\%.

The spatial variations of the LF faint end are also mentioned by
other authors. The work of Yang et al. (2004) reveals that cluster
A168 consists of two merging subclusters; the LF of A168S shows a
decay tail, while the LF of A168N shows an increasing one. They
interpreted it by the cannibalism model of Hausman \& Ostriker
(1978), and regarded the formation and evolution of these two
subclusters might be different. The nearby cluster Coma is also a
good example. Adami et al. (2007) found that there is clearly a
dichotomy between the LF in the north-northeast regions and the LF
in the south-southwest regions. The former has a steeply rising LF
end, but the latter has a much flatter LF end. They believe that the
merging of faint galaxies with brighter objects may partly paly a role in
flattening the faint end slope of the LF. Moreover, by using a sample
of clusters with only two substructures, Krywult (2007) investigated
the LF of galaxies in each substructure. He also found a similar
result with us: the end slope of the LF is lower in the dominant
substructure than in the second one.

\subsection{\sl Star Formation Histories of Member Galaxies}
The star formation histories (SFHs) of galaxies in a cluster
may shed some light on the dynamical evolution of the cluster.
Within our field of view, there are 136 member galaxies that have
been spectroscopically observed by the SDSS, which facilitates us
to investigate the SFHs of galaxies with different morphological
types, and to explore the environmental effect on the SFHs of the
member galaxies.

The spectra of these galaxies are extracted from the SDSS DR6.
We take the STARLIGHT spectral synthesis code (Bruzual \& Charlot 2003)
which aims to decompose an observed spectrum into a series of
simple stellar populations (SSPs) of various ages and metallicities.
Each SSP contributes a fraction $x_{j}$ (j=1,$\cdots$,$N_{\star}$) to
the flux at a chosen normalization wavelength ($=4020\AA$). By
running STARLIGHT, astrophysically interesting output can be obtained,
such as the SFHs of a galaxy, its extinction and velocity dispersion
(Cid Fernandes et al. 2005). A base of $N_{\star}=45$ SSPs were
used from the evolutionary synthesis models of Bruzual \& Charlot (2003),
spanning 15 ages ($t_{\star j}$) between 1 Myr to 13 Gyr and 3
metallicities, $Z=$ 0.2, 1, and 2.5 $Z_{\odot}$. Stellar extinction
was modelled with the extinction law of Cardelli, Clayton \& Mathis (1989)
with $R_V=3.1$. Here, we mainly emphasis on the statistics of the
characteristic stellar ages for all of these member
galaxies. To characterize the SFH of a specific galaxy, the mean
stellar age, which is defined to condense the whole age distribution
of SSPs to a single number, is used (Asari et al. 2007),
\begin{equation}
<logt_{\star}>_{L}=\sum\limits_{j=1}^{N_{\star}}x_{j}log t_{\star,j}
\end{equation}
where the subscript L denotes a light-weighted average.

Many previous works have focused on studying the variation of SFHs
with the Hubble type (Gallagher et al. 1984; Sandage 1986; Abraham
et al. 1999). They suggested that the properties of the
Hubble sequence in past times can be systematized by considering
the time variation of the star-formation rate. In our work, the
morphological types were obtained by our SED-fitting code based on
the HYPERZ, and the classification indices, ranging from 1 to 7,
are defined to denote E, S0, Sa, Sb, Sc, Sd, and Im galaxies,
respectively. Figure \ref{age-type} shows the mean stellar ages of
member galaxies in A1775A (denoted by ``$+$'') and A1775B (denoted
by ``$\bullet$'') along the Hubble sequence. In general, the
light-weighted mean stellar ages of early-type galaxies are found
to be systematically greater than those of late-type galaxies, and
they exhibit a larger dispersion. Among all of the different types
of galaxies, the mean stellar age distribution of S0 galaxies is
very similar to that of E galaxies. For the late-type galaxies, the
more significant is the disk that the galaxy possesses, the younger
is the mean stellar age that it has. It is likely that early-type
galaxies burst their star-formation activity at an early epoch and
deplete the gas rapidly, and no gas disk in such galaxies can be
seen now. Meanwhile, late-type galaxies exhibit on average a more
protracted SFH. Since the stellar mass and dynamical environment
could also influence the SFHs, a certain dispersion in the
distribution of the mean stellar ages is reasonable, even for those
galaxies with the same Hubble type.

It is well accepted that environment plays a key role in the evolution
of galaxies (Gobat et al. 2008; Braglia et al. 2009). Figure \ref{age-dist}
shows how the galaxies' SFHs vary with their distances to the center of
each subcluster, which is defined by the peaks of X-ray surface brightness.
For subcluster A1775B (panel b), the mean stellar ages of galaxies
seem to decrease with the increasing cluster centric distance. A linear
fit to member galaxies of A1775B gives $<logt_{\star}>_{L}=-0.31
(\pm0.09)Radius+9.68(\pm0.12)$, and the correlation coefficient is -0.34.
This trend is partly due to the contribution of the morphology-density
relation (Dressler 1980). The majority of early-type galaxies reside in
the core region of A1775B (with a radius smaller than 1 Mpc), they have
relatively high mean stellar ages, while most late-type galaxies
residing in the outskirts of A1775B have lower mean stellar ages, on
average. To separate between the morphology-density relation and the
SF-density relation, the galaxy sample is divided into two populations,
the early-type (denoted by ``$\bullet$'') and the late-type galaxies
(denoted by ``$\ast$''). It is noticed that the early-type galaxies
present a similar trend to the age gradient shown by the solid line
in Figure \ref{age-dist}b, while the late-type galaxies seem to show
an opposite trend. A physical interpretation of this phenomenon is
beyond the scope of this paper. However, for galaxies in A1775A, no
clear environment effect on the galaxies'SFHs is found, and a large
$<log t_\star>_L$ dispersion can be seen, even for galaxies in the core
region. To further investigate the environmental trend of the late-type
galaxies in a cluster, more member galaxies and a sample of clusters are
needed. It should be mentioned that the fraction of late-type galaxies
in A1775A is about 28\%, more than that in A1775B (19\%), which indicates
that the loose subcluster A1775A is still a dynamically young system.

\section{Summary}
This paper presents the multicolor photometry for the nearby rich
cluster of galaxies A1775, using the 60/90 cm Schmidt Telescope of
the NAOC. A1775 is also covered by SDSS photometry and
spectroscopy. After an aperture correction of the SDSS magnitudes
and a flux calibration of the BATC magnitudes, the SEDs obtained
by these two photometric systems were carefully combined. For a
sample of bright galaxies with known spectroscopic redshifts, by
comparing their $z_{ph}$ with $z_{sp}$, the reliability of the
combined SEDs could be verified. For galaxies without redshift
information, after applying the photometric redshift technique, 146
faint galaxies were selected as probable cluster members. By
adding them to the sample of 151 spectroscopically confirmed
member galaxies (sample I), we achieved an enlarged sample of
297 cluster galaxies (sample II).

Based on the positions and redshifts of member galaxies in A1775,
the spatial distribution and dynamics of the cluster galaxies were
investigated. Spatially, A1775 has a double structure, a
subcluster with a lower redshift, A1775A, is located about $14'$
SE to the main concentration A1775B. These two substructures are
also detected in the X-ray brightness map. The spatial
distribution of the galaxies in sample II makes the substructure
A1775A more significant, and a new clump A1775C is found. The KMM
partition of the galaxies in sample I shows that there are 49 and
102 spectroscopically confirmed member galaxies in A1775A and
A1775B, respectively. Compared with the main concentration A1775B,
A1775A appears to be a poor and loose subcluster with a smaller
velocity dispersion of $\sim 432kms^{-1}$. Two central dominant
elliptical galaxies, one of them associated with the head-tail
radio source, are assigned to different subclusters. The original
dominant galaxy in A1775A seems to be ``pulled out'' by the
gravitation of the main subcluster A1775B.

With the 20-band SEDs of the member galaxies, the LF function for
each subcluster was fitted. Supplementing the faint member galaxies
makes it possible to constrain the LF shape at the faint end. The
remarkable difference in the LFs between the two subclusters indicates
that they are at different stages of dynamical evolution. A1775B is
a more dynamically evolved system, while A1775A is still dynamically
young. In fact, the spatial variations of the LF faint end have also
been noticed by other authors. This phenomena are more common
in clusters with substructures, especially with bimodal structures
(Adami et al. 2007; Yang et al. 2004; Krywult 2007). The similar
results of other people give us confidence in our result. By the
STARLIGHT spectral synthesis code, the SFHs of the member galaxies
with available SDSS spectroscopy have been studied. The variations
of the mean stellar ages with Hubble type have been verified,
in the sense that early-type galaxies are likely to have higher
mean stellar ages. The environmental effect on the mean stellar ages
of cluster galaxies is found in A1775B. The galaxies in the core
region of A1775B are likely to have longer mean stellar ages,
which means that these massive galaxies form the bulk of their
stars earlier than those in the outskirts. Such an environmental
effect is not found for galaxies in the poor subcluster A1775A.
\\

We would like to thank the referee who gives the invaluable
suggestions to improve the paper. Our research is based on data
collected at National Astronomical Observatory of China, which
is operated by Chinese Academy of Sciences. We are grateful to
all staff members for their support during observations.
The database SDSS and NED are also used in our work. We would
like to thank Prof. Weihao Bian, Mr. Wei Jing at the Nanjing
Normal university, Prof. Jun Ma, Jianghua Wu, and Zhenyu Wu at
the National Astronomical Observatories and Prof. Kong Xu,
Dr. Lin Lin at the University of Science and Technology of
China for stimulating discussion and helpful suggestions.
This work was supported by the National Natural Science
Foundation of China (NSFC) under Nos.10778618 and 10633020,
and by the National Basic Research Program of China (973
Program) under No.2007CB815403.


\newpage

\begin{figure}
  \begin{center}
    \FigureFile(85mm,80mm){f1.eps}
  \end{center}
\caption{Distribution of the zero-points between the SDSS and BATC
photometric systems for 2248 galaxies bright than $r'=21.5$ mag.}
\label{zero-dis}
\end{figure}

\begin{figure}
  \begin{center}
  \includegraphics[width=76mm]{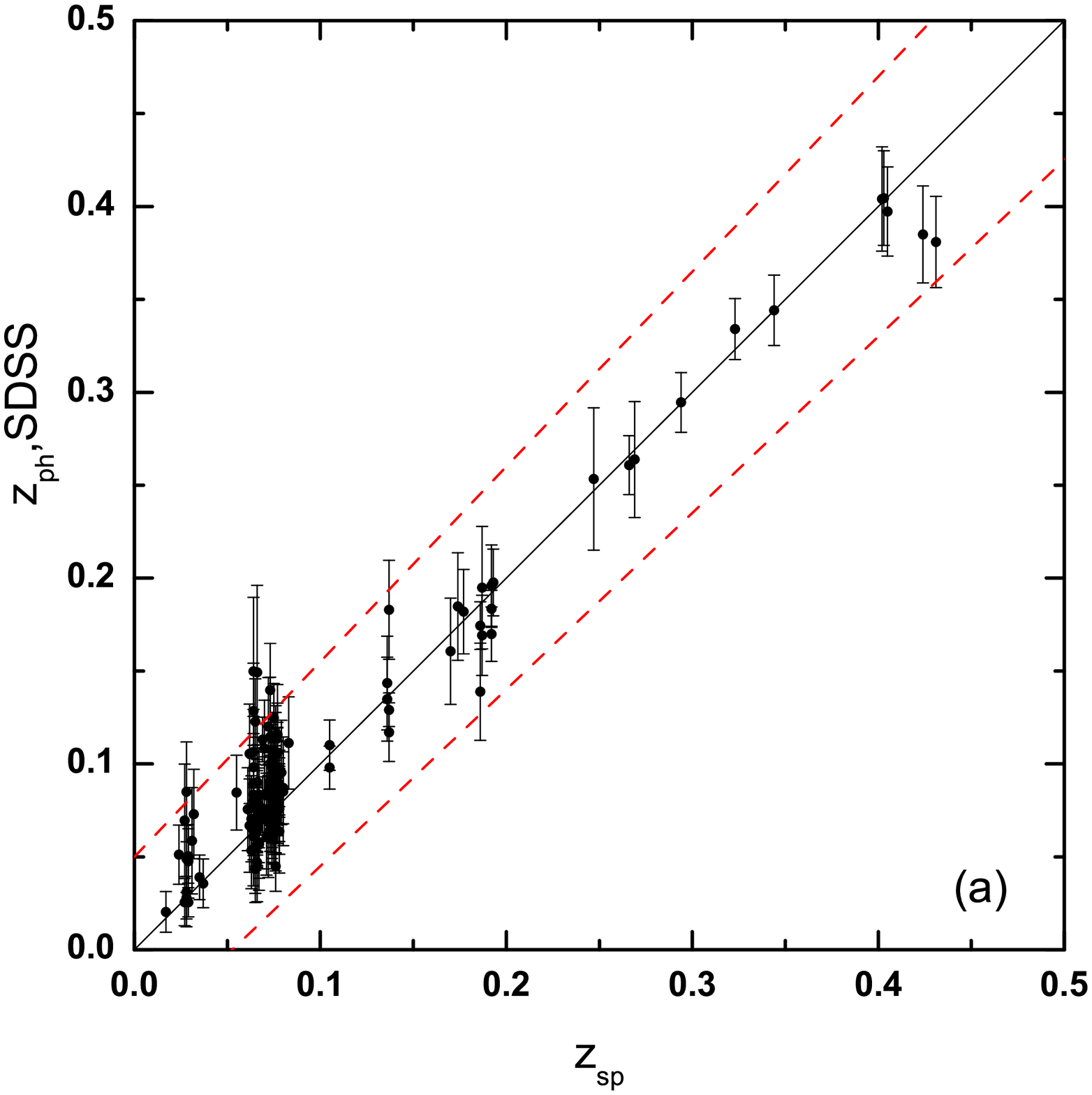}
  \includegraphics[width=80mm]{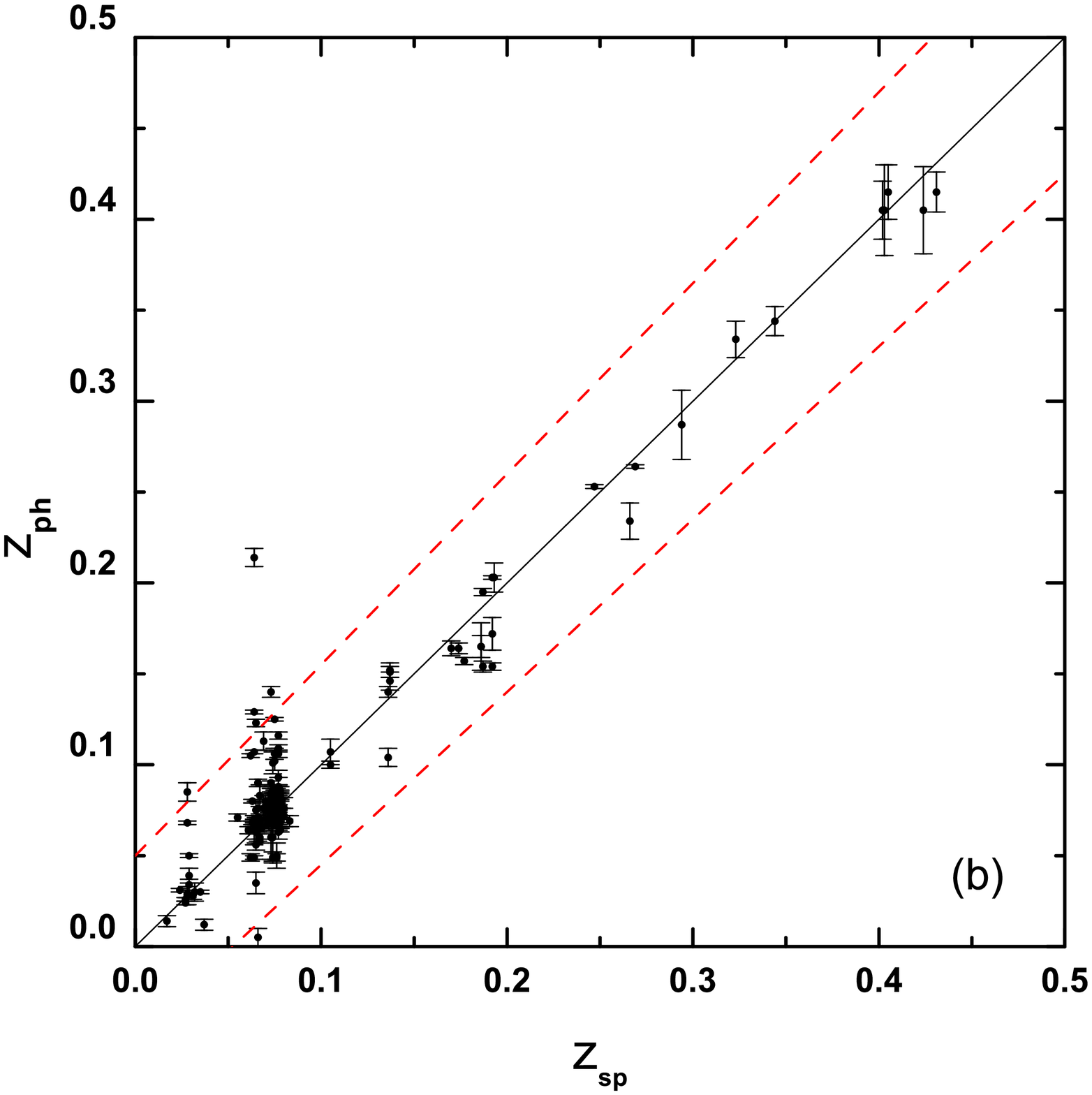}
  \end{center}
\caption{Comparison between $z_{ph}$ and $z_{sp}$ for 190 galaxies
with known spectroscopic redshifts in the region of A1775: (a)~$z_{ph}$~
derived by 5 broad-band photometric data from SDSS. (b)~$z_{ph}$~derived
by the combined SDSS and BATC SEDs.}
\label{z-z}
\end{figure}

\begin{figure}
  \begin{center}
  \includegraphics[width=82mm]{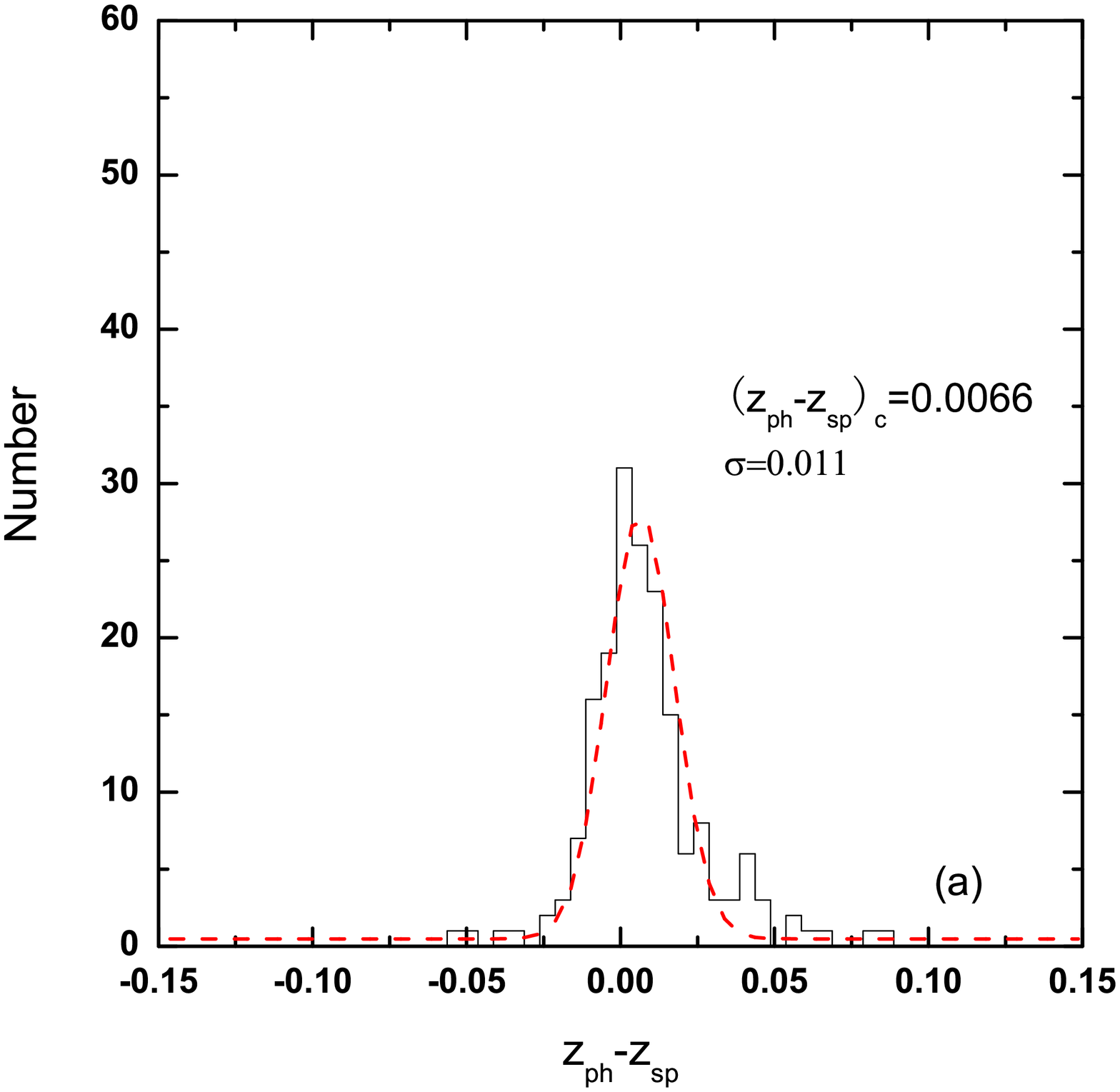}
  \includegraphics[width=80mm]{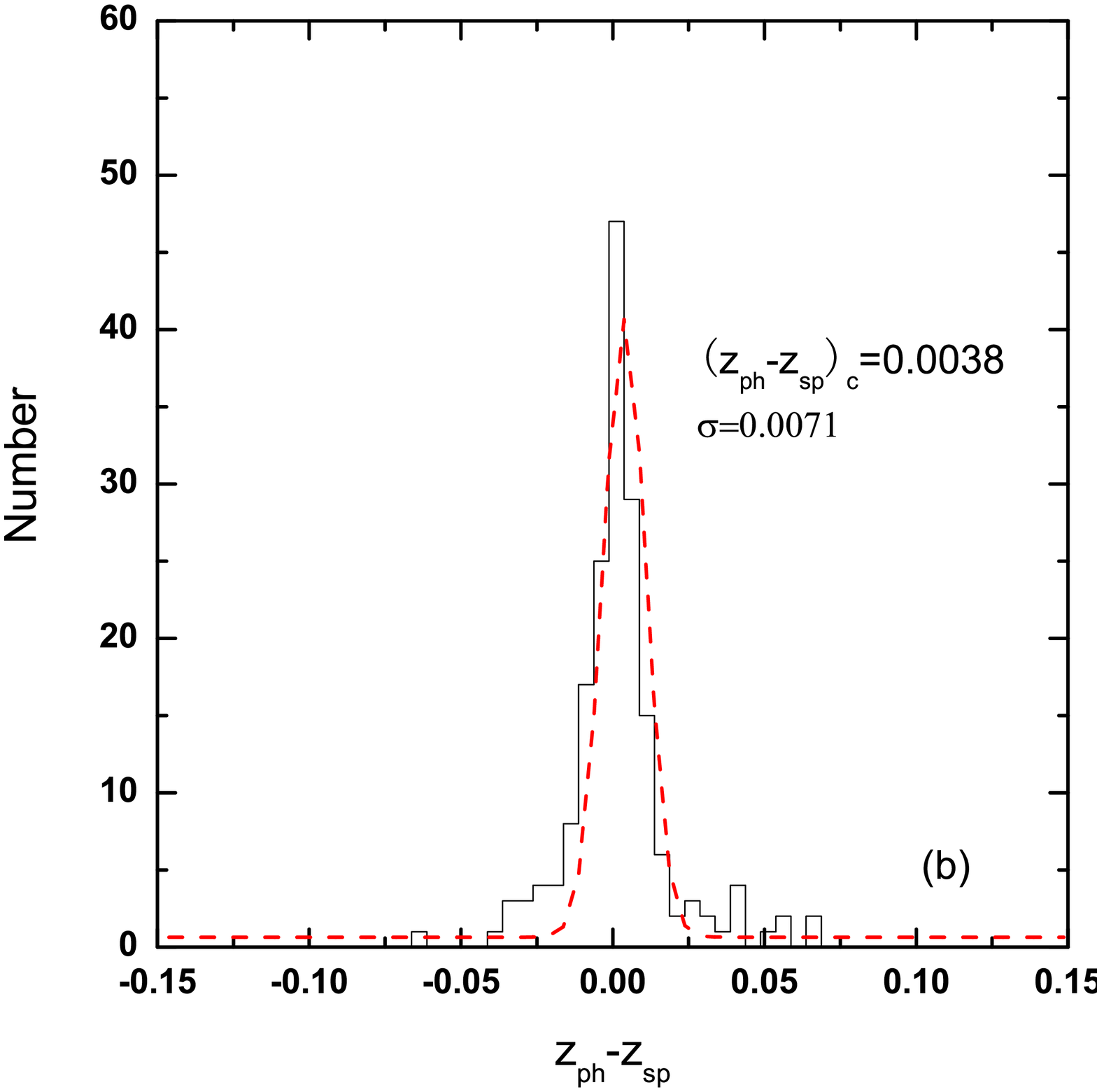}
  \end{center}
\caption{The distribution of offsets~$z_{ph}-z_{sp}$~for 190 bright
galaxies. (a)~$z_{ph}$~derived by 5 broad bands photometric data from SDSS.
(b)~$z_{ph}$~derived by the combined SDSS and BATC SEDs. The dashed lines
present Gaussian fittings.}
\label{zz-dis}
\end{figure}

\begin{figure}
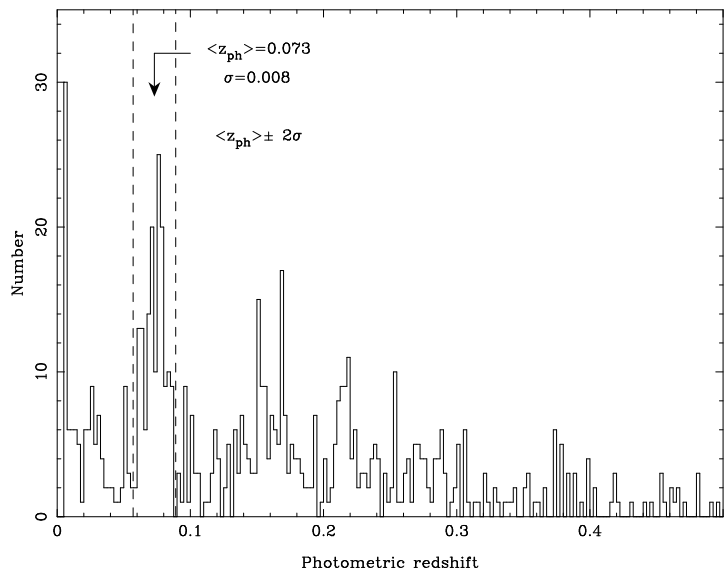

  \begin{center}
    \FigureFile(95mm,80mm){f4.eps}
  \end{center}
\caption{Distribution of $z_{ph}$ for 657
galaxies detected by photometries only.}
\label{zph-cand}
\end{figure}

\begin{figure}
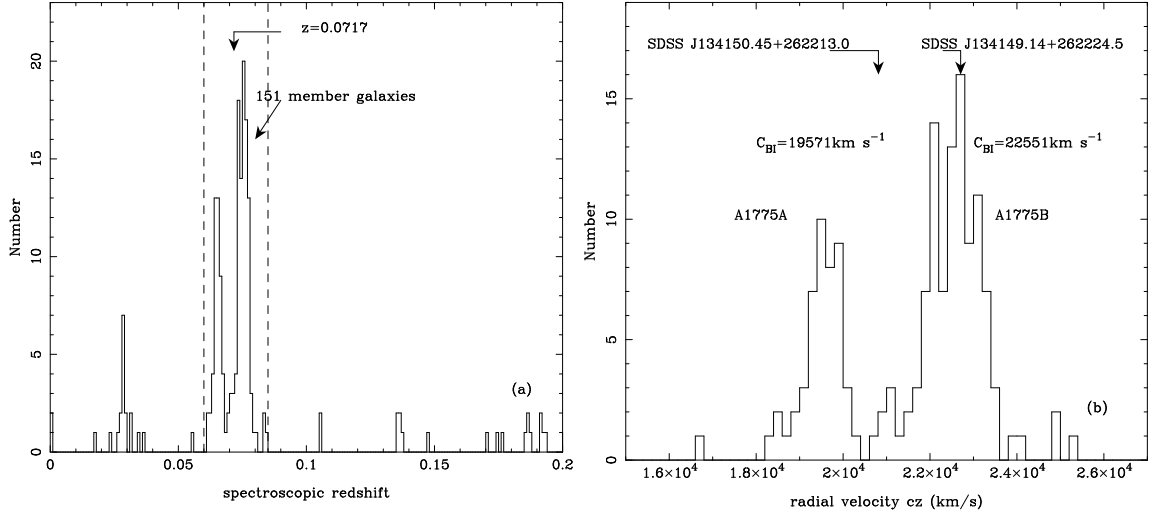

  \begin{center}
    \FigureFile(75mm,80mm){f5-1.eps}
    \FigureFile(75mm,80mm){f5-2.eps}
  \end{center}
\caption{(a) Distribution of spectroscopic redshifts for 224 galaxies. The bin size is 0.001. (b)~Histogram of observed velocities for galaxies in the left panel. The bin size is 200 km$s^{-1}$. The velocities of the two central dominant galaxies are denoted by arrows.}
\label{cz}
\end{figure}

\begin{figure}
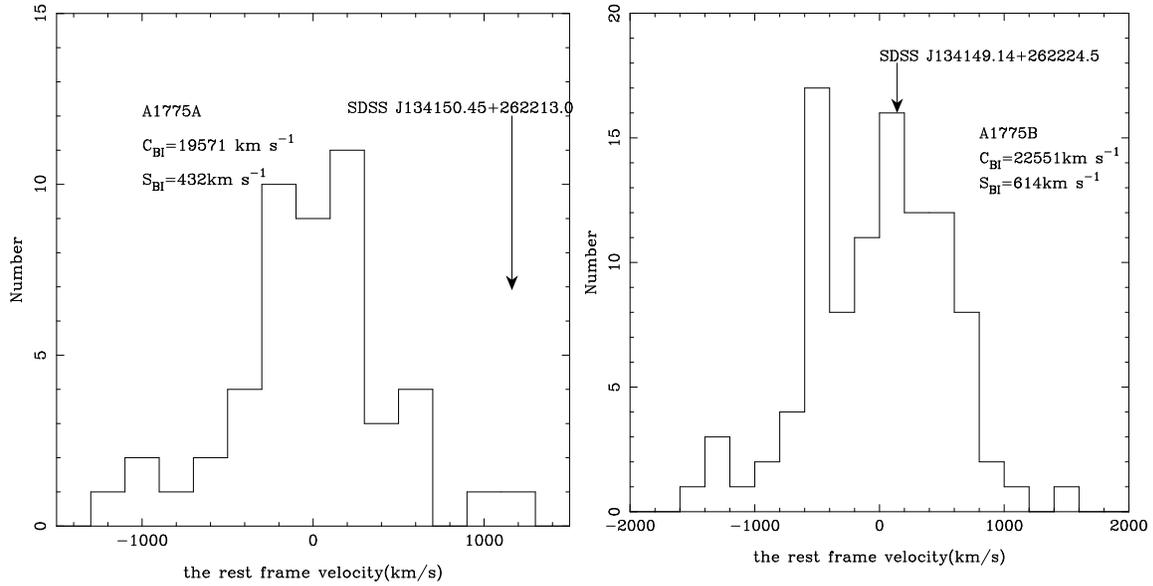

  \begin{center}
    \FigureFile(75mm,75mm){f6-1.eps}
    \FigureFile(75mm,75mm){f6-2.eps}
  \end{center}
\caption{(Left) Rest-frame velocity distribution of galaxies in A1775A.
(Right) Rest-frame velocity distribution of galaxies in A1775B. The arrows
mark the velocities of the two central dominant galaxies.}
\label{rest-cz}
\end{figure}

\begin{figure}
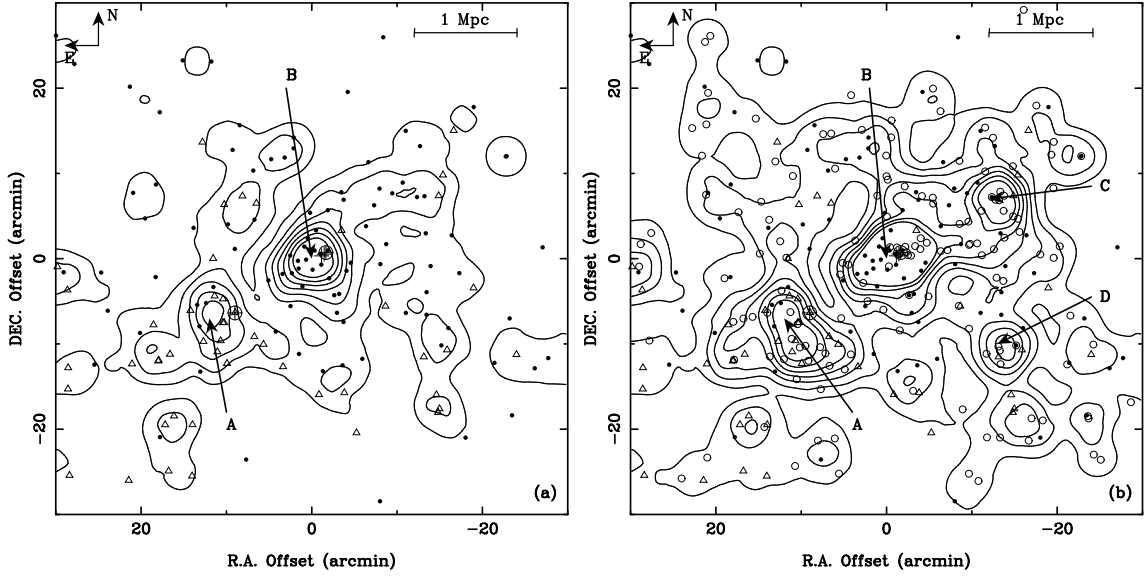

  \begin{center}
    \FigureFile(75mm,75mm){f7-1.eps}
    \FigureFile(75mm,75mm){f7-2.eps}
  \end{center}
\caption{(a) Spatial distribution of 151 spectroscopically confirmed
member galaxies, including 49 member galaxies of A1775A (denoted by
``$\bigtriangleup$'') and 102 member galaxies of A1775B (denoted by
``$\bullet$''); (b) Spatial distribution of enlarged sample of member
galaxies. The newly selected member galaxies are denoted by ``$\circ$''.
For both panels, the contour maps of the surface density for these
galaxies, using the smoothing window with a radius of $2'$, are
superposed. The contour levels are 0.05, 0.10, 0.15, 0.20, 0.25, 0.30
and 0.35 $arcmin^{-2}$, respectively. The positions for two central
giant elliptical galaxies, are denoted by ``$+$'', and the peaks of
X-ray emission are denoted by ``$\oplus$''.}
\label{surfdence}
\end{figure}

\begin{figure}
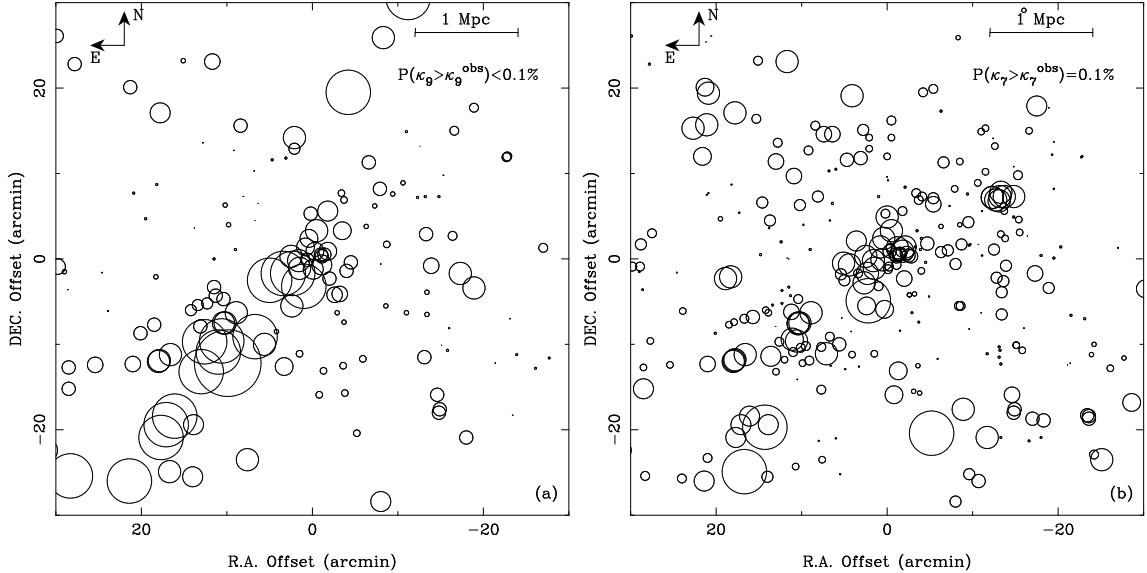

  \begin{center}
    \FigureFile(75mm,75mm){f8-1.eps}
    \FigureFile(75mm,75mm){f8-2.eps}
  \end{center}
\caption{(a)~Bubble plots showing the localized variation for (a)~sample I
(with neighbor size n=9), and (b)~sample II (with neighbor size n=7).}
\label{bubble}
\end{figure}

\begin{figure}
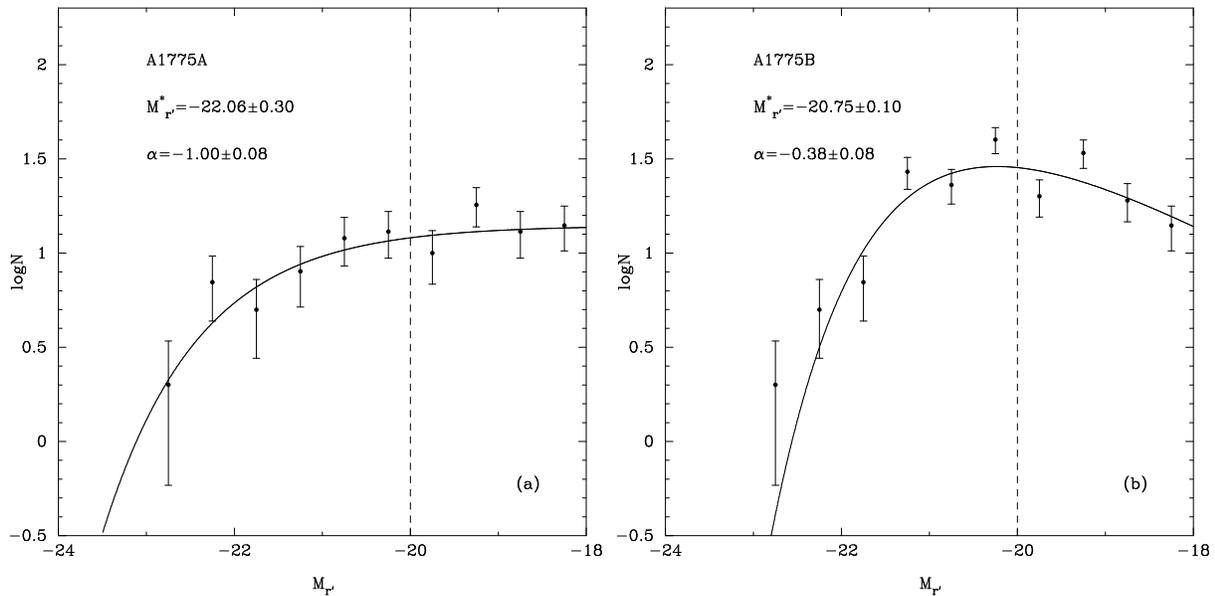

  \begin{center}
    \FigureFile(160mm,80mm){f9.eps}
  \end{center}
\caption{Luminosity functions for galaxies in subclusters
(a) A1775A, and (b) A1775B. The dashed line corresponds to the
limiting magnitude of the SDSS spectral observation.}
\label{lf-AB}
\end{figure}

\begin{figure}
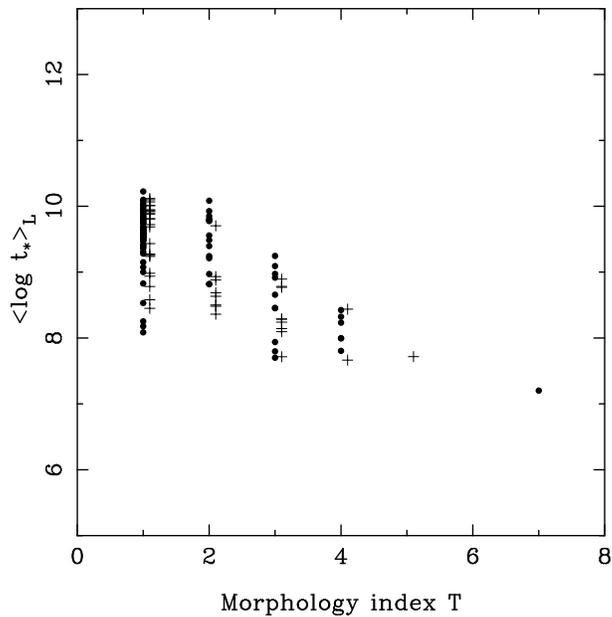

  \begin{center}
    \FigureFile(80mm,80mm){f10.eps}
  \end{center}
\caption{Mean stellar ages as a function of morphological indices
T, which ranges from 1 to 7, corresponding to E, S0, Sa, Sb, Sc, Sd, and
Im galaxies, respectively. The member galaxies in A1775A are denoted by
``$+$'', and those in A1775B are denoted by ``$\bullet$''.}
\label{age-type}
\end{figure}

\begin{figure}
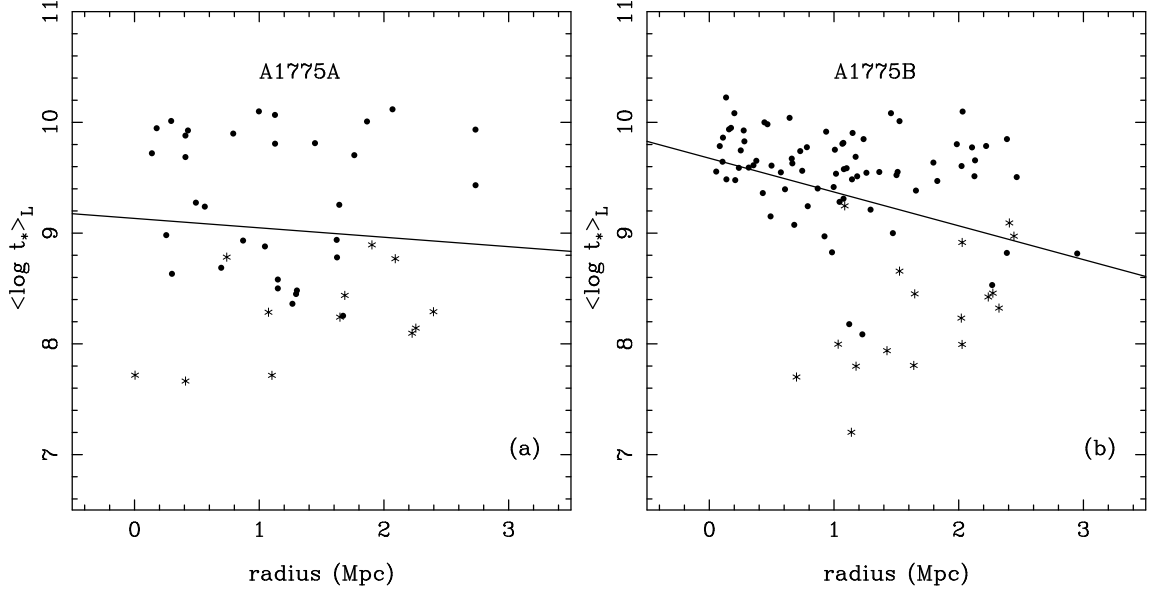

  \begin{center}
    \FigureFile(75mm,75mm){f11-1.eps}
    \FigureFile(75mm,75mm){f11-2.eps}
  \end{center}
\caption{Mean stellar ages of member galaxies as functions of
the cluster centric radii of substructures (a) A1775A, and (b) A1775B.
Early-type galaxies (including E, S0) are denoted by ``$\bullet$'',
late-type galaxies (including Sa, Sb, Sc, Sd, and Im) are denoted by ``$\ast$''.
The solid line represents a linear fit to all member galaxies of each subcluster.}
\label{age-dist}
\end{figure}

\def\baselinestretch{1.2}
\begin{table}[h]
\caption{Details of the BATC filters and our observations}
\label{table1}
\footnotesize
\begin{center}
\begin{tabular}{ccccccccc}   \hline
\noalign{\smallskip}
 No. & Filter & $\lambda_{c}$ & FWHM &
Exposure &  Number of & Seeing  & Objects & Completeness \\
   &  name & (\AA) & (\AA) &
(second) & Images     & (arcsec)  & Detected & magnitude \\
\noalign{\smallskip}   \hline \noalign{\smallskip}
  1  & a  & 3360 & 222 & 16800  &14&  4.45   & 4411 & 21.5 \\
  2  & b  & 3890 & 291 & 15600  &13&  4.57   & 4660 & 21.0 \\
  3  & c  & 4210 & 309 & 9000   &10&  4.11   & 5281 & 20.5 \\
  4  & d  & 4550 & 332 & 18000  &15&  4.47   & 4743 & 20.5 \\
  5  & e  & 4920 & 374 & 13200  &11&  7.25   & 4962 & 20.5 \\
  6  & f  & 5270 & 344 & 10800  & 9&  5.46   & 4818 & 20.0 \\
  7  & g  & 5795 & 289 & 7200   & 6&  3.88   & 5145 & 20.0 \\
  8  & h  & 6075 & 308 & 8400   & 7&  4.46   & 5237 & 20.0 \\
  9  & i  & 6660 & 491 & 8400   & 7&  4.17   & 5181 & 19.5 \\
  10 & j  & 7050 & 238 & 6000   & 5&  4.36   & 5179 & 19.5 \\
  11 & k  & 7490 & 192 & 7200   & 6&  4.12   & 5214 & 19.0 \\
  12 & m  & 8020 & 255 & 10800  & 9&  5.20   & 4972 & 19.0 \\
  13 & n  & 8480 & 167 & 8400   & 7&  5.27   & 4735 & 19.0 \\
  14 & o  & 9190 & 247 & 14400  &12&  4.94   & 4444 & 18.5 \\
  15 & p  & 9745 & 275 & 19200  &16&  4.75   & 4381 & 18.5 \\
\noalign{\smallskip}   \hline
\end{tabular}
\end{center}
\end{table}

\begin{table}
\caption{Catalog of 146 photometric-selected candidates of member galaxies in A1775}
\vspace{1mm}
\label{table2}
\def\baselinestretch{1.2}
\centering
\vspace{5mm} \tiny
 \tabcolsep 0.4mm
\begin{tabular}{rcccccccccccccccccccccccc}
\hline\noalign{\smallskip} {No.}&{R.A.} &{Decl.} & {$z_{\rm ph}$}
& {$T$} & {$a$} & {$b$} & {$c$} & {$d$} & {$e$} & {$f$} & {$g$} &
{$h$} & {$i$} & {$j$} & {$k$} & {$m$}
& {$n$} & {$o$} & {$p$}& {$u'$} &{$g'$} &{$r'$} &{$i'$}&{$z'$} \\
\noalign{\smallskip}\hline\noalign{\smallskip}
    1  & 205.0162 & 25.97159  & 0.080 & 1 & 19.61 & 19.16 & 18.66 & 18.45 & 18.46 & 18.18 & 18.21 & 18.10 & 18.05 & 18.23 & 18.11 & 17.99 & 17.96 & 99.00 & 99.00 & 19.23 & 18.41 & 18.17 & 18.06 & 18.1\\
    2  & 205.0313 & 25.98172  & 0.061 & 3 & 21.51 & 21.30 & 20.88 & 20.28 & 19.15 & 19.68 & 18.90 & 19.16 & 19.06 & 18.88 & 19.17 & 18.52 & 18.26 & 18.66 & 18.33 & 20.89 & 19.97 & 19.27 & 19.00 & 18.8\\
    3  & 205.0315 & 26.19678  & 0.087 & 1 & 20.78 & 19.79 & 19.37 & 18.95 & 18.44 & 18.17 & 17.87 & 17.82 & 17.56 & 17.51 & 17.46 & 17.38 & 17.25 & 17.24 & 17.09 & 20.23 & 18.59 & 17.81 & 17.44 & 17.1\\
    4  & 205.0425 & 26.05245  & 0.075 & 1 & 20.46 & 19.71 & 19.43 & 19.08 & 18.99 & 18.95 & 18.84 & 18.95 & 18.82 & 18.44 & 18.69 & 18.55 & 18.85 & 99.00 & 99.00 & 20.07 & 19.14 & 18.87 & 18.72 & 18.6\\
    5  & 205.0438 & 26.05689  & 0.084 & 1 & 20.44 & 19.79 & 19.38 & 19.04 & 18.80 & 18.69 & 18.58 & 18.47 & 18.34 & 18.05 & 99.00 & 17.98 & 17.92 & 99.00 & 17.77 & 19.65 & 18.69 & 18.36 & 18.10 & 17.9\\
    6  & 205.0565 & 26.56520  & 0.075 & 1 & 19.74 & 18.69 & 18.29 & 17.57 & 17.17 & 16.92 & 16.61 & 16.53 & 16.30 & 16.20 & 16.13 & 15.98 & 15.94 & 15.87 & 15.72 & 19.31 & 17.43 & 16.49 & 16.13 & 15.8\\
    7  & 205.0813 & 26.25348  & 0.060 & 1 & 20.66 & 20.22 & 19.55 & 19.26 & 18.69 & 18.58 & 18.45 & 18.14 & 18.07 & 18.04 & 17.99 & 17.77 & 17.60 & 17.64 & 17.63 & 20.95 & 18.99 & 18.23 & 17.90 & 17.6\\
    8  & 205.0886 & 26.15800  & 0.079 & 1 & 21.61 & 20.41 & 19.72 & 19.06 & 18.90 & 18.52 & 18.38 & 18.18 & 17.97 & 17.90 & 17.74 & 17.60 & 17.62 & 17.65 & 17.59 & 20.75 & 18.98 & 18.15 & 17.75 & 17.5\\
    9  & 205.1040 & 26.38880  & 0.059 & 1 & 21.53 & 21.15 & 20.58 & 20.28 & 19.81 & 19.60 & 19.19 & 19.09 & 19.13 & 18.99 & 18.87 & 18.81 & 18.99 & 18.69 & 19.12 & 21.65 & 19.91 & 19.15 & 18.84 & 18.6\\
   10  & 205.1165 & 26.59306  & 0.060 & 3 & 22.22 & 20.92 & 20.48 & 20.13 & 18.43 & 19.43 & 19.66 & 19.24 & 19.34 & 19.07 & 19.38 & 18.75 & 18.82 & 18.78 & 18.59 & 21.06 & 20.22 & 19.48 & 19.12 & 18.9\\
   11  & 205.1173 & 26.36645  & 0.078 & 3 & 21.14 & 20.15 & 19.88 & 19.36 & 18.71 & 18.62 & 18.31 & 18.23 & 18.13 & 18.11 & 18.05 & 17.86 & 17.74 & 17.86 & 17.44 & 20.65 & 19.15 & 18.32 & 17.96 & 17.6\\
   12  & 205.1175 & 26.50228  & 0.087 & 1 & 22.11 & 21.13 & 21.24 & 20.73 & 19.67 & 19.89 & 19.15 & 19.29 & 19.07 & 18.67 & 19.10 & 19.03 & 19.32 & 18.59 & 19.02 & 22.19 & 20.09 & 19.33 & 18.96 & 18.7\\
   13  & 205.1218 & 26.56800  & 0.069 & 2 & 20.84 & 20.40 & 19.64 & 19.20 & 18.78 & 18.52 & 18.29 & 18.13 & 17.96 & 17.89 & 17.83 & 17.56 & 17.61 & 17.53 & 17.54 & 20.72 & 19.02 & 18.16 & 17.78 & 17.4\\
   14  & 205.1420 & 26.04852  & 0.068 & 7 & 20.55 & 20.51 & 19.74 & 19.37 & 19.25 & 19.28 & 19.19 & 18.90 & 18.94 & 18.98 & 19.12 & 18.70 & 18.72 & 99.00 & 18.75 & 20.56 & 19.58 & 19.14 & 18.87 & 18.8\\
   15  & 205.1465 & 26.37294  & 0.078 & 4 & 20.41 & 19.98 & 19.84 & 19.24 & 19.11 & 18.87 & 18.81 & 18.69 & 18.53 & 18.40 & 18.36 & 18.10 & 18.35 & 18.33 & 17.74 & 20.33 & 19.36 & 18.69 & 18.34 & 18.1\\
   16  & 205.1543 & 26.66405  & 0.070 & 7 & 20.88 & 20.12 & 19.65 & 19.80 & 19.66 & 19.32 & 18.90 & 19.13 & 18.98 & 18.94 & 19.21 & 18.86 & 18.66 & 19.17 & 18.97 & 20.57 & 19.70 & 19.28 & 19.00 & 18.9\\
   17  & 205.1663 & 26.05261  & 0.075 & 2 & 20.35 & 20.22 & 19.52 & 19.13 & 18.91 & 18.60 & 18.29 & 18.28 & 18.16 & 18.00 & 18.08 & 17.83 & 17.82 & 99.00 & 17.59 & 20.41 & 19.05 & 18.33 & 18.00 & 17.8\\
   18  & 205.1727 & 26.00896  & 0.069 & 5 & 21.36 & 20.00 & 19.34 & 19.26 & 19.06 & 18.71 & 18.43 & 18.51 & 18.33 & 99.00 & 18.31 & 18.11 & 18.14 & 99.00 & 99.00 & 20.13 & 19.01 & 18.54 & 18.25 & 18.0\\
   19  & 205.1814 & 26.85108  & 0.071 & 1 & 14.27 & 13.17 & 12.21 & 87.57 & 87.57 & 12.03 & 87.57 & 11.74 & 11.78 & 11.48 & 87.57 & 87.57 & 87.57 & 87.57 & 87.57 & 20.53 & 19.53 & 19.29 & 19.08 & 18.9\\
   20  & 205.1916 & 26.41702  & 0.065 & 1 & 21.38 & 20.64 & 20.52 & 19.80 & 19.30 & 19.39 & 19.12 & 19.15 & 19.08 & 18.82 & 19.27 & 18.60 & 18.67 & 18.64 & 18.58 & 21.44 & 19.86 & 19.32 & 19.03 & 18.8\\
   21  & 205.1946 & 26.44264  & 0.074 & 1 & 20.79 & 20.98 & 19.72 & 19.01 & 18.77 & 18.51 & 18.17 & 18.03 & 17.82 & 17.77 & 17.63 & 17.49 & 17.49 & 17.40 & 17.20 & 21.06 & 18.85 & 18.00 & 17.64 & 17.4\\
   22  & 205.1995 & 26.19530  & 0.060 & 1 & 18.27 & 17.63 & 17.07 & 16.65 & 16.19 & 16.14 & 15.92 & 15.78 & 15.58 & 15.44 & 15.43 & 15.28 & 15.20 & 15.18 & 15.00 & 18.41 & 16.57 & 15.82 & 15.36 & 15.1\\
   23  & 205.2056 & 26.44757  & 0.065 & 2 & 20.84 & 21.65 & 19.75 & 19.37 & 18.69 & 18.67 & 18.59 & 18.22 & 18.04 & 18.04 & 17.97 & 17.67 & 17.57 & 17.56 & 17.30 & 20.99 & 19.10 & 18.28 & 17.89 & 17.6\\
   24  & 205.2058 & 26.54549  & 0.075 & 3 & 21.96 & 20.49 & 20.40 & 20.07 & 19.21 & 19.16 & 19.01 & 18.76 & 18.46 & 18.54 & 18.37 & 18.08 & 18.31 & 17.95 & 17.70 & 21.39 & 19.52 & 18.75 & 18.35 & 18.1\\
   25  & 205.2246 & 26.40583  & 0.072 & 3 & 21.39 & 22.11 & 20.54 & 21.34 & 20.47 & 19.85 & 19.18 & 19.57 & 19.04 & 19.44 & 18.96 & 18.91 & 19.01 & 18.89 & 17.92 & 21.57 & 20.17 & 19.42 & 19.09 & 18.8\\
   26  & 205.2250 & 26.48781  & 0.079 & 3 & 21.10 & 99.00 & 19.69 & 19.18 & 18.86 & 18.71 & 18.51 & 18.39 & 18.19 & 18.14 & 18.10 & 17.96 & 17.87 & 17.83 & 17.95 & 21.21 & 19.32 & 18.45 & 18.00 & 17.7\\
   27  & 205.2263 & 26.45820  & 0.071 & 2 & 20.39 & 21.72 & 19.58 & 19.24 & 18.82 & 18.62 & 18.38 & 18.35 & 18.11 & 18.00 & 18.16 & 17.76 & 17.89 & 17.82 & 17.43 & 20.70 & 19.11 & 18.38 & 18.06 & 17.8\\
   28  & 205.2272 & 26.32115  & 0.080 & 4 & 20.60 & 99.00 & 20.10 & 19.55 & 19.25 & 19.14 & 18.84 & 18.77 & 18.76 & 18.50 & 18.56 & 18.41 & 18.51 & 18.51 & 17.87 & 20.85 & 19.49 & 18.84 & 18.49 & 18.3\\
   29  & 205.2334 & 26.49573  & 0.083 & 1 & 21.73 & 22.25 & 20.75 & 19.98 & 19.25 & 19.33 & 19.18 & 19.11 & 18.80 & 18.76 & 18.73 & 18.55 & 18.46 & 18.42 & 18.70 & 21.74 & 19.90 & 19.10 & 18.69 & 18.5\\
   30  & 205.2336 & 26.15857  & 0.083 & 3 & 20.11 & 19.95 & 19.18 & 18.76 & 18.45 & 18.30 & 18.09 & 17.94 & 17.67 & 17.60 & 17.58 & 17.33 & 17.25 & 17.17 & 17.10 & 20.16 & 18.78 & 17.98 & 17.50 & 17.2\\
   31  & 205.2350 & 26.01515  & 0.087 & 3 & 20.77 & 20.19 & 19.99 & 19.28 & 19.38 & 19.17 & 18.74 & 18.66 & 18.58 & 18.60 & 18.38 & 18.35 & 18.41 & 18.31 & 18.02 & 20.75 & 19.44 & 18.72 & 18.33 & 18.0\\
   32  & 205.2353 & 26.18516  & 0.067 & 1 & 21.57 & 20.39 & 19.52 & 19.35 & 18.98 & 18.92 & 18.38 & 18.48 & 18.31 & 18.19 & 18.30 & 18.09 & 18.15 & 18.01 & 17.80 & 20.56 & 19.19 & 18.55 & 18.18 & 18.0\\
   33  & 205.2361 & 26.47894  & 0.087 & 3 & 21.97 & 99.00 & 20.52 & 20.08 & 19.27 & 19.23 & 19.06 & 18.82 & 18.56 & 18.71 & 18.47 & 18.29 & 18.25 & 18.38 & 18.34 & 21.46 & 19.71 & 18.86 & 18.42 & 18.1\\
   34  & 205.2382 & 26.19302  & 0.075 & 4 & 20.39 & 20.38 & 19.34 & 19.24 & 19.02 & 18.64 & 18.64 & 18.58 & 18.36 & 18.12 & 18.41 & 18.14 & 18.10 & 18.17 & 17.98 & 20.55 & 19.19 & 18.64 & 18.27 & 18.1\\
   35  & 205.2419 & 26.32740  & 0.070 & 1 & 19.84 & 21.62 & 18.70 & 18.47 & 18.48 & 18.23 & 18.23 & 18.09 & 18.06 & 17.81 & 18.19 & 18.01 & 17.97 & 17.94 & 17.80 & 19.72 & 18.58 & 18.21 & 18.02 & 17.8\\
   36  & 205.2439 & 26.47992  & 0.083 & 1 & 19.97 & 99.00 & 19.05 & 18.89 & 18.85 & 18.44 & 18.30 & 18.32 & 18.17 & 18.01 & 18.18 & 18.12 & 18.09 & 18.29 & 18.36 & 19.67 & 18.66 & 18.34 & 18.11 & 18.0\\
   37  & 205.2475 & 26.38325  & 0.075 & 1 & 20.40 & 23.42 & 19.70 & 19.18 & 18.97 & 18.85 & 18.62 & 18.56 & 18.37 & 18.32 & 18.16 & 18.11 & 18.25 & 18.06 & 17.54 & 20.87 & 19.29 & 18.52 & 18.18 & 17.9\\
   38  & 205.2491 & 26.60114  & 0.071 & 1 & 20.71 & 20.25 & 19.14 & 18.74 & 18.21 & 18.15 & 17.88 & 17.76 & 17.57 & 17.52 & 17.46 & 17.32 & 17.22 & 17.40 & 17.20 & 20.23 & 18.54 & 17.78 & 17.41 & 17.2\\
   39  & 205.2521 & 26.48535  & 0.071 & 2 & 18.79 & 99.00 & 17.70 & 17.26 & 17.05 & 16.87 & 16.62 & 16.53 & 16.35 & 16.21 & 16.20 & 16.07 & 16.03 & 15.97 & 15.78 & 18.66 & 17.25 & 16.53 & 16.17 & 15.9\\
   40  & 205.2633 & 26.01538  & 0.063 & 4 & 20.88 & 20.67 & 20.04 & 19.95 & 19.75 & 19.63 & 19.27 & 19.46 & 19.26 & 19.08 & 19.00 & 18.96 & 18.81 & 99.00 & 18.43 & 20.86 & 19.78 & 19.29 & 19.14 & 18.9\\
   41  & 205.2641 & 26.17893  & 0.068 & 1 & 19.87 & 20.33 & 18.89 & 18.91 & 18.52 & 18.40 & 18.25 & 18.22 & 18.04 & 18.01 & 18.11 & 17.88 & 17.98 & 17.82 & 17.52 & 20.02 & 18.74 & 18.29 & 18.01 & 17.9\\
   42  & 205.2662 & 26.53539  & 0.069 & 4 & 21.90 & 20.89 & 20.08 & 20.10 & 19.61 & 19.32 & 19.33 & 19.37 & 19.17 & 18.88 & 18.97 & 18.59 & 18.34 & 18.65 & 18.20 & 20.73 & 19.87 & 19.33 & 18.97 & 18.7\\
   43  & 205.2663 & 26.62112  & 0.068 & 1 & 21.16 & 21.22 & 19.93 & 19.61 & 19.40 & 19.02 & 18.61 & 18.48 & 18.45 & 18.37 & 18.21 & 18.04 & 18.06 & 17.97 & 17.80 & 21.25 & 19.38 & 18.54 & 18.17 & 17.9\\
   44  & 205.2737 & 26.37158  & 0.079 & 4 & 20.74 & 99.00 & 20.73 & 20.37 & 19.40 & 19.84 & 19.52 & 19.10 & 19.21 & 19.25 & 19.26 & 18.84 & 19.64 & 18.72 & 18.55 & 21.19 & 19.95 & 19.36 & 19.02 & 18.8\\
   45  & 205.2745 & 26.45824  & 0.065 & 1 & 21.67 & 99.00 & 20.02 & 19.48 & 19.22 & 18.78 & 18.48 & 18.34 & 18.24 & 18.23 & 18.26 & 17.83 & 17.84 & 17.92 & 17.78 & 21.51 & 19.31 & 18.48 & 18.10 & 17.7\\
   46  & 205.2820 & 25.93016  & 0.060 & 1 & 21.73 & 20.58 & 20.17 & 19.76 & 19.52 & 19.27 & 19.11 & 19.07 & 19.02 & 18.99 & 18.71 & 18.77 & 18.58 & 99.00 & 99.00 & 21.13 & 19.72 & 19.04 & 18.76 & 18.6\\
   47  & 205.2841 & 26.39458  & 0.078 & 1 & 20.64 & 99.00 & 19.75 & 19.38 & 19.13 & 19.16 & 19.09 & 18.95 & 18.86 & 18.94 & 18.80 & 18.64 & 19.20 & 19.17 & 19.22 & 20.39 & 19.28 & 18.97 & 18.79 & 18.6\\
   48  & 205.2995 & 26.39366  & 0.070 & 3 & 22.00 & 99.00 & 20.68 & 20.00 & 20.22 & 19.71 & 19.54 & 19.58 & 19.35 & 19.19 & 19.02 & 18.83 & 19.18 & 18.66 & 18.34 & 21.55 & 20.18 & 19.45 & 19.14 & 18.9\\
   49  & 205.3017 & 26.39169  & 0.060 & 3 & 21.45 & 99.00 & 19.53 & 18.95 & 18.49 & 18.43 & 18.17 & 18.00 & 17.87 & 17.78 & 17.68 & 17.51 & 17.43 & 17.31 & 17.14 & 20.57 & 18.87 & 18.03 & 17.63 & 17.3\\
   50  & 205.3021 & 26.22793  & 0.079 & 1 & 21.72 & 99.00 & 20.48 & 19.45 & 20.13 & 19.44 & 19.05 & 19.12 & 18.98 & 19.19 & 18.71 & 18.59 & 18.61 & 18.75 & 18.92 & 21.57 & 19.91 & 19.10 & 18.76 & 18.6\\
   51  & 205.3026 & 25.94435  & 0.060 & 1 & 20.63 & 20.52 & 19.99 & 19.58 & 19.71 & 19.39 & 19.25 & 19.02 & 19.23 & 19.20 & 19.30 & 18.96 & 18.79 & 99.00 & 99.00 & 21.04 & 19.67 & 19.26 & 19.08 & 19.0\\
   52  & 205.3032 & 26.43705  & 0.066 & 1 & 20.67 & 99.00 & 19.71 & 19.55 & 19.27 & 19.37 & 19.75 & 19.34 & 19.29 & 18.92 & 19.10 & 18.98 & 18.64 & 20.56 & 18.30 & 20.75 & 19.56 & 19.27 & 19.15 & 19.0\\
   53  & 205.3161 & 26.07020  & 0.070 & 6 & 20.71 & 20.44 & 19.71 & 19.62 & 19.38 & 19.31 & 18.99 & 18.97 & 18.86 & 18.66 & 18.65 & 18.52 & 18.55 & 18.76 & 19.30 & 20.36 & 19.38 & 18.95 & 18.66 & 18.4\\
   54  & 205.3224 & 26.43257  & 0.060 & 4 & 20.51 & 19.49 & 19.75 & 19.34 & 17.86 & 18.64 & 18.76 & 18.40 & 18.62 & 18.59 & 18.49 & 17.89 & 17.87 & 17.95 & 17.92 & 21.30 & 19.54 & 18.67 & 18.28 & 17.9\\
   55  & 205.3226 & 26.55598  & 0.069 & 2 & 20.75 & 99.00 & 19.79 & 19.13 & 18.76 & 18.69 & 18.38 & 18.38 & 18.04 & 18.02 & 17.95 & 17.56 & 17.70 & 17.59 & 17.59 & 21.07 & 19.14 & 18.33 & 17.96 & 17.6\\
   56  & 205.3243 & 26.27225  & 0.062 & 3 & 20.37 & 19.20 & 19.08 & 18.61 & 18.07 & 18.02 & 17.79 & 17.63 & 17.46 & 17.38 & 17.32 & 17.07 & 16.99 & 16.91 & 16.76 & 19.83 & 18.42 & 17.67 & 17.25 & 16.9\\
   57  & 205.3489 & 26.38148  & 0.068 & 1 & 21.24 & 20.14 & 19.72 & 19.22 & 18.68 & 18.59 & 18.29 & 18.24 & 18.01 & 17.83 & 17.83 & 17.70 & 17.64 & 17.54 & 17.36 & 20.88 & 19.09 & 18.21 & 17.83 & 17.5\\
   58  & 205.3633 & 26.37946  & 0.063 & 7 & 20.68 & 19.66 & 19.39 & 19.10 & 18.91 & 18.96 & 18.92 & 18.61 & 18.60 & 18.48 & 18.95 & 18.31 & 18.68 & 18.68 & 17.99 & 20.07 & 19.19 & 18.85 & 18.65 & 18.6\\
   59  & 205.3637 & 26.65299  & 0.077 & 1 & 21.37 & 20.86 & 19.85 & 19.32 & 18.59 & 18.59 & 18.32 & 18.22 & 18.00 & 17.95 & 17.87 & 17.75 & 17.58 & 17.54 & 17.23 & 20.87 & 19.07 & 18.23 & 17.85 & 17.5\\
   60  & 205.3798 & 26.47230  & 0.075 & 1 & 20.81 & 19.74 & 19.41 & 18.66 & 18.14 & 18.10 & 17.82 & 17.74 & 17.51 & 17.42 & 17.41 & 17.22 & 17.20 & 17.13 & 16.81 & 20.43 & 18.55 & 17.75 & 17.36 & 17.0\\
   61  & 205.3800 & 26.69726  & 0.067 & 1 & 20.82 & 20.52 & 19.51 & 19.11 & 18.53 & 18.50 & 18.10 & 18.02 & 17.86 & 17.71 & 17.85 & 17.53 & 17.44 & 17.51 & 17.18 & 20.63 & 18.88 & 18.10 & 17.73 & 17.4\\
   62  & 205.3807 & 26.48401  & 0.075 & 2 & 20.69 & 20.10 & 19.73 & 19.26 & 18.79 & 18.55 & 18.17 & 18.10 & 17.92 & 17.92 & 17.72 & 17.59 & 17.60 & 17.46 & 17.28 & 20.65 & 18.91 & 18.10 & 17.72 & 17.4\\
   63  & 205.3904 & 26.29108  & 0.059 & 1 & 21.06 & 19.98 & 19.52 & 18.92 & 18.29 & 18.31 & 18.03 & 17.82 & 17.67 & 17.57 & 17.49 & 17.38 & 17.28 & 17.20 & 17.25 & 20.69 & 18.74 & 17.84 & 17.46 & 17.1\\
   64  & 205.3940 & 26.39590  & 0.086 & 1 & 21.22 & 20.29 & 20.07 & 19.90 & 19.05 & 18.87 & 18.83 & 18.45 & 18.32 & 18.14 & 18.31 & 18.05 & 17.98 & 18.08 & 17.67 & 21.17 & 19.41 & 18.55 & 18.17 & 17.9\\
   65  & 205.4045 & 26.48711  & 0.071 & 1 & 20.05 & 19.24 & 19.01 & 18.49 & 17.91 & 17.72 & 17.51 & 17.34 & 17.14 & 17.05 & 17.04 & 16.84 & 16.82 & 16.74 & 16.50 & 20.16 & 18.23 & 17.35 & 16.99 & 16.7\\
   66  & 205.4072 & 26.40472  & 0.064 & 1 & 20.67 & 19.98 & 19.61 & 19.21 & 18.63 & 18.43 & 18.18 & 17.99 & 17.86 & 17.82 & 17.73 & 17.53 & 17.48 & 17.38 & 17.11 & 21.05 & 18.91 & 18.06 & 17.65 & 17.3\\
   67  & 205.4081 & 26.38361  & 0.065 & 3 & 20.61 & 20.09 & 19.70 & 19.18 & 18.65 & 18.46 & 18.31 & 18.20 & 17.92 & 17.96 & 17.88 & 17.66 & 17.33 & 17.41 & 17.27 & 20.86 & 19.07 & 18.22 & 17.84 & 17.5\\
   68  & 205.4194 & 26.38936  & 0.076 & 2 & 20.56 & 19.98 & 20.05 & 19.26 & 18.76 & 18.58 & 18.38 & 18.24 & 17.93 & 17.86 & 17.80 & 17.57 & 17.46 & 17.50 & 17.40 & 20.86 & 19.09 & 18.18 & 17.81 & 17.4\\
   69  & 205.4238 & 26.10641  & 0.070 & 1 & 20.71 & 21.75 & 19.73 & 19.38 & 19.34 & 19.26 & 19.01 & 19.14 & 18.78 & 18.86 & 18.84 & 18.71 & 18.69 & 18.52 & 18.62 & 20.95 & 19.51 & 19.04 & 18.88 & 18.6\\
   70  & 205.4283 & 26.37076  & 0.064 & 1 & 21.70 & 20.27 & 20.01 & 19.43 & 18.76 & 18.67 & 18.29 & 18.19 & 18.02 & 17.95 & 18.00 & 17.79 & 17.71 & 17.58 & 17.43 & 21.00 & 19.12 & 18.27 & 17.88 & 17.5\\
   71  & 205.4325 & 26.29340  & 0.079 & 1 & 19.83 & 18.88 & 18.51 & 18.02 & 17.58 & 17.46 & 17.23 & 17.14 & 16.97 & 16.93 & 16.82 & 16.75 & 16.67 & 16.63 & 16.59 & 19.47 & 17.86 & 17.13 & 16.82 & 16.5\\
   72  & 205.4351 & 26.37163  & 0.074 & 3 & 21.16 & 20.02 & 19.97 & 19.62 & 18.93 & 18.70 & 18.37 & 18.41 & 18.07 & 18.35 & 18.02 & 17.78 & 17.75 & 17.65 & 17.34 & 20.99 & 19.28 & 18.41 & 18.01 & 17.7\\
   73  & 205.4379 & 26.37739  & 0.074 & 1 & 20.72 & 19.87 & 19.71 & 18.77 & 18.62 & 18.34 & 18.22 & 17.89 & 17.74 & 17.75 & 17.63 & 17.40 & 17.44 & 17.35 & 17.32 & 20.72 & 18.81 & 17.94 & 17.57 & 17.2\\
   \noalign{\smallskip}\hline
\end{tabular}
\end{table}

\begin{table}
\setcounter{table}{1}
\def\baselinestretch{1.2}
 \centering

\begin{minipage}{35mm}
 \caption{\it --- Continued.}
\end{minipage}

\tiny 
\tabcolsep 0.4mm
\begin{tabular}{rcccccccccccccccccccccccc}
\hline\noalign{\smallskip} {No.}&{R.A.} &{Decl.} & {$z_{\rm ph}$}
& {$T$} & {$a$} & {$b$} & {$c$} & {$d$} & {$e$} & {$f$} & {$g$} &
{$h$} & {$i$} & {$j$} & {$k$} & {$m$}
& {$n$} & {$o$} & {$p$}& {$u'$} &{$g'$} &{$r'$} &{$i'$}&{$z'$} \\
\noalign{\smallskip}\hline\noalign{\smallskip}
   74  & 205.4409 & 26.38803  & 0.079 & 1 & 20.20 & 19.10 & 18.92 & 18.20 & 17.79 & 17.55 & 17.30 & 17.18 & 16.94 & 16.88 & 16.76 & 16.63 & 16.55 & 16.49 & 16.23 & 20.11 & 18.06 & 17.16 & 16.76 & 16.4\\
   75  & 205.4421 & 26.50595  & 0.064 & 2 & 21.50 & 20.42 & 20.27 & 19.88 & 19.30 & 19.40 & 19.22 & 18.86 & 18.68 & 18.52 & 18.70 & 18.33 & 18.34 & 18.30 & 17.88 & 21.32 & 19.69 & 18.93 & 18.58 & 18.3\\
   76  & 205.4523 & 26.37495  & 0.075 & 1 & 22.54 & 21.34 & 20.91 & 20.16 & 19.46 & 19.40 & 19.14 & 18.98 & 18.80 & 18.71 & 18.59 & 18.39 & 18.29 & 18.25 & 18.12 & 21.56 & 19.81 & 19.07 & 18.49 & 18.1\\
   77  & 205.4529 & 26.22454  & 0.060 & 3 & 21.21 & 20.17 & 19.87 & 19.55 & 18.95 & 18.98 & 18.71 & 18.52 & 18.39 & 18.38 & 18.42 & 18.11 & 18.03 & 17.90 & 18.18 & 21.16 & 19.45 & 18.65 & 18.27 & 17.9\\
   78  & 205.4548 & 26.37348  & 0.074 & 1 & 18.77 & 17.67 & 17.38 & 16.63 & 16.06 & 15.84 & 15.56 & 15.42 & 15.19 & 15.12 & 14.99 & 14.85 & 14.76 & 14.69 & 14.51 & 18.92 & 16.34 & 15.41 & 14.98 & 14.6\\
   79  & 205.4589 & 26.38669  & 0.079 & 2 & 21.17 & 20.02 & 19.44 & 19.01 & 18.51 & 18.13 & 17.80 & 17.73 & 17.50 & 17.36 & 17.30 & 17.21 & 17.15 & 17.06 & 16.90 & 20.56 & 18.64 & 17.72 & 17.29 & 16.9\\
   80  & 205.4706 & 26.63488  & 0.075 & 2 & 21.17 & 99.00 & 20.08 & 19.67 & 19.27 & 18.99 & 18.84 & 18.90 & 18.59 & 18.47 & 18.48 & 18.20 & 18.29 & 18.04 & 18.32 & 20.91 & 19.51 & 18.82 & 18.51 & 18.2\\
   81  & 205.4715 & 26.60178  & 0.071 & 3 & 20.82 & 99.00 & 20.55 & 20.29 & 20.39 & 19.67 & 19.76 & 19.56 & 19.13 & 19.10 & 19.13 & 18.77 & 18.70 & 18.88 & 18.20 & 21.77 & 20.24 & 19.47 & 19.14 & 18.8\\
   82  & 205.4740 & 26.38728  & 0.075 & 2 & 21.10 & 20.44 & 20.02 & 19.35 & 19.28 & 18.79 & 18.43 & 18.34 & 18.07 & 17.95 & 17.76 & 17.64 & 17.75 & 17.33 & 17.21 & 20.98 & 19.14 & 18.24 & 17.84 & 17.5\\
   83  & 205.4766 & 26.34972  & 0.076 & 2 & 20.63 & 19.82 & 19.50 & 19.02 & 18.44 & 18.19 & 17.92 & 17.85 & 17.54 & 17.50 & 17.42 & 17.21 & 17.11 & 17.09 & 16.90 & 20.47 & 18.71 & 17.83 & 17.41 & 17.1\\
   84  & 205.4790 & 26.51913  & 0.071 & 5 & 19.93 & 19.26 & 19.00 & 18.75 & 18.76 & 18.41 & 18.29 & 18.26 & 18.07 & 17.72 & 17.98 & 17.83 & 17.72 & 18.06 & 18.38 & 19.82 & 18.68 & 18.23 & 17.93 & 17.8\\
   85  & 205.4807 & 26.44679  & 0.074 & 1 & 19.31 & 18.51 & 18.19 & 17.74 & 17.31 & 17.15 & 16.91 & 16.79 & 16.62 & 16.49 & 16.46 & 16.32 & 16.29 & 16.19 & 16.02 & 19.07 & 17.55 & 16.79 & 16.44 & 16.1\\
   86  & 205.4818 & 26.52594  & 0.073 & 3 & 20.73 & 19.90 & 19.84 & 19.34 & 18.82 & 18.75 & 18.45 & 18.33 & 18.15 & 18.05 & 17.97 & 17.72 & 17.60 & 17.71 & 17.57 & 20.87 & 19.17 & 18.34 & 17.95 & 17.7\\
   87  & 205.4823 & 26.56624  & 0.068 & 3 & 20.38 & 19.53 & 19.47 & 19.04 & 18.47 & 18.48 & 18.08 & 18.18 & 17.92 & 17.89 & 17.81 & 17.66 & 17.63 & 17.49 & 17.46 & 20.54 & 18.89 & 18.17 & 17.79 & 17.4\\
   88  & 205.4885 & 26.26500  & 0.084 & 2 & 20.65 & 19.71 & 19.59 & 18.81 & 18.42 & 18.28 & 18.01 & 17.86 & 17.64 & 17.60 & 17.47 & 17.33 & 17.17 & 17.12 & 17.03 & 20.37 & 18.71 & 17.81 & 17.51 & 17.1\\
   89  & 205.5017 & 26.48785  & 0.079 & 1 & 20.29 & 19.32 & 19.04 & 18.33 & 17.87 & 17.56 & 17.27 & 17.20 & 16.98 & 16.90 & 16.77 & 16.66 & 16.55 & 16.55 & 16.35 & 20.09 & 18.12 & 17.17 & 16.78 & 16.5\\
   90  & 205.5230 & 26.28170  & 0.076 & 2 & 20.63 & 19.50 & 19.34 & 18.81 & 18.40 & 18.06 & 17.76 & 17.63 & 17.44 & 17.40 & 17.31 & 17.11 & 17.14 & 17.05 & 16.84 & 20.33 & 18.52 & 17.65 & 17.27 & 16.9\\
   91  & 205.5323 & 26.31730  & 0.064 & 1 & 21.26 & 20.27 & 20.08 & 19.45 & 19.23 & 19.00 & 18.61 & 18.47 & 18.34 & 18.06 & 18.10 & 18.04 & 18.13 & 17.85 & 17.75 & 20.93 & 19.30 & 18.44 & 18.10 & 17.8\\
   92  & 205.5356 & 26.61642  & 0.079 & 2 & 20.63 & 99.00 & 19.75 & 19.27 & 18.95 & 18.60 & 18.36 & 18.23 & 17.97 & 17.81 & 17.85 & 17.60 & 17.53 & 17.31 & 17.23 & 21.97 & 19.24 & 18.28 & 17.76 & 17.4\\
   93  & 205.5500 & 26.40002  & 0.075 & 2 & 21.80 & 20.56 & 20.16 & 20.33 & 19.27 & 19.18 & 19.08 & 18.79 & 18.46 & 18.32 & 18.35 & 18.26 & 18.44 & 18.20 & 17.67 & 21.33 & 19.59 & 18.73 & 18.36 & 18.0\\
   94  & 205.5590 & 26.68434  & 0.083 & 1 & 20.35 & 20.71 & 19.11 & 18.79 & 18.84 & 18.49 & 18.38 & 18.36 & 18.16 & 17.84 & 18.09 & 17.97 & 18.07 & 17.96 & 17.80 & 20.02 & 18.82 & 18.34 & 18.05 & 18.0\\
   95  & 205.5593 & 26.17764  & 0.073 & 3 & 20.18 & 99.00 & 19.46 & 19.20 & 18.93 & 18.64 & 18.27 & 18.15 & 17.98 & 17.89 & 17.89 & 17.48 & 17.70 & 17.45 & 17.46 & 20.11 & 19.04 & 18.23 & 17.78 & 17.4\\
   96  & 205.5646 & 26.35302  & 0.073 & 1 & 20.63 & 19.91 & 19.67 & 19.41 & 18.85 & 18.69 & 18.57 & 18.33 & 18.14 & 18.00 & 17.96 & 17.86 & 17.86 & 17.76 & 17.42 & 20.87 & 19.09 & 18.30 & 17.96 & 17.7\\
   97  & 205.5727 & 26.14832  & 0.087 & 1 & 20.88 & 99.00 & 20.09 & 20.07 & 19.24 & 19.00 & 18.72 & 18.61 & 18.53 & 18.39 & 18.36 & 18.30 & 18.15 & 18.38 & 18.71 & 21.34 & 19.41 & 18.67 & 18.29 & 18.0\\
   98  & 205.5773 & 26.35624  & 0.086 & 1 & 21.50 & 19.87 & 19.81 & 19.31 & 19.00 & 18.64 & 18.42 & 18.33 & 18.18 & 18.13 & 17.98 & 17.98 & 17.81 & 17.83 & 17.62 & 20.89 & 19.14 & 18.33 & 17.97 & 17.6\\
   99  & 205.5804 & 26.24850  & 0.084 & 1 & 22.11 & 20.03 & 20.52 & 19.99 & 19.46 & 19.33 & 19.24 & 18.85 & 18.74 & 18.51 & 18.55 & 18.39 & 18.43 & 18.50 & 18.55 & 22.96 & 19.68 & 18.86 & 18.53 & 18.2\\
  100  & 205.5828 & 26.33373  & 0.087 & 3 & 21.21 & 19.93 & 20.09 & 19.16 & 18.87 & 18.65 & 18.45 & 18.27 & 18.09 & 18.05 & 17.96 & 17.79 & 17.81 & 17.56 & 17.34 & 20.75 & 19.19 & 18.31 & 17.90 & 17.5\\
  101  & 205.5939 & 26.54071  & 0.076 & 1 & 20.49 & 19.64 & 19.44 & 19.03 & 18.59 & 18.44 & 18.24 & 18.07 & 17.95 & 17.88 & 17.79 & 17.67 & 17.62 & 17.65 & 17.43 & 20.45 & 18.81 & 18.08 & 17.78 & 17.5\\
  102  & 205.5942 & 26.01339  & 0.074 & 3 & 21.10 & 20.32 & 20.03 & 99.00 & 19.15 & 18.89 & 99.00 & 18.87 & 18.34 & 18.11 & 18.09 & 17.90 & 18.00 & 99.00 & 17.75 & 20.97 & 19.35 & 18.60 & 18.27 & 17.9\\
  103  & 205.6026 & 26.60943  & 0.070 & 1 & 20.46 & 99.00 & 19.38 & 19.42 & 19.03 & 18.98 & 18.80 & 18.99 & 18.58 & 18.42 & 18.69 & 18.44 & 19.27 & 18.58 & 19.93 & 20.55 & 19.28 & 18.88 & 18.69 & 18.5\\
  104  & 205.6057 & 26.15535  & 0.064 & 1 & 20.41 & 99.00 & 19.61 & 19.09 & 18.67 & 18.46 & 18.13 & 18.13 & 17.88 & 17.87 & 17.86 & 17.58 & 17.37 & 17.42 & 17.30 & 20.82 & 18.95 & 18.16 & 17.80 & 17.5\\
  105  & 205.6147 & 26.17929  & 0.075 & 1 & 21.06 & 99.00 & 20.21 & 19.78 & 19.63 & 19.33 & 18.87 & 19.27 & 18.87 & 18.75 & 18.78 & 18.70 & 18.84 & 18.55 & 17.99 & 21.57 & 19.92 & 19.16 & 18.82 & 18.6\\
  106  & 205.6199 & 26.60827  & 0.080 & 7 & 20.30 & 99.00 & 19.64 & 19.47 & 19.30 & 19.19 & 18.90 & 19.08 & 18.86 & 19.11 & 18.61 & 18.51 & 18.74 & 18.60 & 18.86 & 20.17 & 19.44 & 19.02 & 18.69 & 18.5\\
  107  & 205.6248 & 26.10972  & 0.075 & 2 & 20.75 & 20.90 & 19.50 & 18.95 & 18.45 & 18.44 & 99.00 & 18.12 & 18.00 & 18.06 & 17.79 & 17.66 & 17.56 & 17.53 & 17.31 & 20.74 & 18.93 & 18.15 & 17.78 & 17.5\\
  108  & 205.6262 & 26.19171  & 0.060 & 1 & 21.25 & 99.00 & 20.38 & 19.87 & 19.52 & 19.45 & 19.32 & 19.26 & 19.02 & 18.94 & 19.14 & 18.73 & 18.63 & 18.65 & 18.41 & 21.88 & 19.98 & 19.35 & 18.99 & 18.8\\
  109  & 205.6307 & 26.00921  & 0.069 & 1 & 19.68 & 18.89 & 18.30 & 17.78 & 17.29 & 17.07 & 16.77 & 16.67 & 16.48 & 16.41 & 16.32 & 16.14 & 16.11 & 16.05 & 15.89 & 19.46 & 17.54 & 16.67 & 16.29 & 16.0\\
  110  & 205.6516 & 26.16515  & 0.075 & 1 & 19.97 & 99.00 & 18.65 & 18.19 & 17.86 & 17.68 & 17.42 & 99.00 & 17.15 & 17.09 & 17.01 & 16.88 & 16.90 & 16.73 & 16.68 & 19.71 & 18.04 & 17.34 & 16.99 & 16.7\\
  111  & 205.6582 & 26.19440  & 0.060 & 1 & 20.00 & 99.00 & 18.97 & 18.73 & 18.62 & 18.51 & 18.26 & 18.31 & 18.24 & 18.15 & 18.04 & 18.00 & 18.18 & 17.87 & 17.66 & 19.88 & 18.70 & 18.28 & 18.08 & 17.9\\
  112  & 205.6705 & 26.19127  & 0.079 & 1 & 20.98 & 99.00 & 20.09 & 19.45 & 19.31 & 19.16 & 99.00 & 19.12 & 19.26 & 19.32 & 18.99 & 18.94 & 19.24 & 18.97 & 18.40 & 20.79 & 19.60 & 19.23 & 18.94 & 18.9\\
  113  & 205.6752 & 26.23673  & 0.068 & 2 & 20.47 & 99.00 & 19.14 & 18.56 & 17.98 & 17.92 & 17.66 & 17.51 & 17.34 & 17.25 & 17.16 & 16.91 & 16.88 & 16.81 & 16.67 & 19.77 & 18.46 & 17.57 & 17.09 & 16.8\\
  114  & 205.6770 & 26.12908  & 0.084 & 1 & 21.33 & 20.52 & 20.15 & 19.40 & 19.27 & 18.89 & 18.80 & 18.42 & 18.30 & 18.42 & 18.14 & 18.05 & 17.90 & 99.00 & 17.89 & 21.22 & 19.26 & 18.46 & 18.12 & 17.8\\
  115  & 205.6804 & 25.95945  & 0.082 & 1 & 21.49 & 20.82 & 19.95 & 19.68 & 19.37 & 99.00 & 19.31 & 19.01 & 19.11 & 18.66 & 18.95 & 18.91 & 18.85 & 99.00 & 99.00 & 20.72 & 19.50 & 19.10 & 18.84 & 18.8\\
  116  & 205.6861 & 26.52641  & 0.073 & 3 & 20.48 & 99.00 & 19.90 & 19.51 & 19.17 & 18.97 & 18.62 & 18.70 & 18.56 & 18.38 & 18.44 & 18.14 & 18.32 & 18.46 & 18.09 & 21.59 & 19.42 & 18.76 & 18.35 & 18.2\\
  117  & 205.6883 & 26.20802  & 0.070 & 2 & 21.28 & 22.12 & 19.76 & 19.26 & 18.96 & 18.93 & 18.72 & 18.72 & 18.55 & 18.25 & 18.51 & 18.26 & 18.38 & 18.46 & 18.17 & 21.21 & 19.39 & 18.76 & 18.40 & 18.3\\
  118  & 205.6911 & 26.26032  & 0.079 & 3 & 21.78 & 99.00 & 19.96 & 19.40 & 19.16 & 18.89 & 18.65 & 18.49 & 18.32 & 18.28 & 18.23 & 18.04 & 18.10 & 17.92 & 17.91 & 20.77 & 19.31 & 18.55 & 18.15 & 17.9\\
  119  & 205.7000 & 26.36656  & 0.075 & 1 & 20.07 & 99.00 & 18.59 & 17.86 & 17.40 & 17.23 & 16.94 & 16.81 & 16.59 & 16.50 & 16.37 & 16.24 & 16.20 & 16.14 & 15.82 & 19.65 & 17.70 & 16.79 & 16.38 & 16.0\\
  120  & 205.7050 & 26.16807  & 0.079 & 2 & 20.78 & 21.02 & 19.81 & 19.70 & 18.90 & 18.82 & 19.17 & 18.86 & 18.71 & 18.47 & 18.61 & 18.25 & 18.34 & 18.20 & 18.22 & 20.49 & 19.34 & 18.92 & 18.52 & 18.9\\
  121  & 205.7107 & 26.30239  & 0.068 & 2 & 19.11 & 99.00 & 18.04 & 17.74 & 17.49 & 17.40 & 17.16 & 17.12 & 16.94 & 16.83 & 16.87 & 16.76 & 16.69 & 16.68 & 16.47 & 19.01 & 17.71 & 17.14 & 16.83 & 16.6\\
  122  & 205.7135 & 26.39865  & 0.071 & 3 & 22.88 & 99.00 & 20.39 & 19.62 & 19.25 & 19.46 & 19.19 & 19.01 & 18.75 & 18.69 & 18.57 & 18.30 & 18.38 & 18.58 & 17.99 & 21.26 & 19.75 & 18.97 & 18.59 & 18.2\\
  123  & 205.7244 & 26.55473  & 0.070 & 6 & 19.54 & 20.26 & 18.73 & 18.54 & 18.27 & 18.21 & 18.06 & 18.03 & 17.88 & 17.64 & 17.75 & 17.67 & 17.62 & 17.44 & 17.46 & 19.50 & 18.46 & 18.05 & 17.69 & 17.5\\
  124  & 205.7357 & 26.17346  & 0.067 & 1 & 20.81 & 20.60 & 19.64 & 19.45 & 19.22 & 18.91 & 18.93 & 18.58 & 18.64 & 18.52 & 18.51 & 18.29 & 18.51 & 18.35 & 17.95 & 20.68 & 19.27 & 18.70 & 18.40 & 18.1\\
  125  & 205.7375 & 26.44071  & 0.075 & 1 & 20.51 & 99.00 & 19.56 & 19.07 & 18.60 & 18.35 & 18.20 & 18.00 & 17.82 & 17.84 & 17.69 & 17.50 & 17.35 & 17.39 & 17.29 & 20.55 & 18.79 & 18.01 & 17.66 & 17.4\\
  126  & 205.7413 & 26.25884  & 0.075 & 3 & 20.73 & 22.35 & 19.65 & 19.26 & 18.80 & 18.66 & 18.43 & 18.33 & 18.08 & 17.99 & 17.94 & 17.78 & 17.74 & 17.82 & 17.50 & 20.51 & 19.13 & 18.35 & 17.90 & 17.6\\
  127  & 205.7472 & 26.03562  & 0.064 & 4 & 19.96 & 19.40 & 18.96 & 18.70 & 18.28 & 18.20 & 99.00 & 17.78 & 17.67 & 17.53 & 17.56 & 17.34 & 17.38 & 99.00 & 17.05 & 19.99 & 18.48 & 17.84 & 17.49 & 17.2\\
  128  & 205.7551 & 26.47604  & 0.070 & 3 & 20.18 & 20.38 & 19.33 & 19.15 & 18.79 & 19.01 & 19.06 & 18.72 & 18.63 & 18.28 & 18.66 & 18.59 & 18.47 & 18.66 & 19.87 & 20.21 & 19.15 & 18.80 & 18.52 & 18.4\\
  129  & 205.7580 & 26.37458  & 0.079 & 3 & 20.44 & 22.04 & 19.85 & 19.17 & 18.92 & 18.67 & 18.22 & 18.27 & 18.02 & 17.97 & 17.74 & 17.60 & 18.00 & 17.64 & 17.34 & 20.68 & 19.16 & 18.22 & 17.79 & 17.5\\
  130  & 205.7674 & 26.63857  & 0.075 & 4 & 21.09 & 20.33 & 19.77 & 19.56 & 19.52 & 19.08 & 18.96 & 18.74 & 18.56 & 18.47 & 18.43 & 18.31 & 18.35 & 18.56 & 18.73 & 20.71 & 19.36 & 18.75 & 18.41 & 18.1\\
  131  & 205.7735 & 26.25107  & 0.079 & 4 & 20.65 & 21.20 & 20.17 & 20.10 & 20.06 & 19.53 & 19.26 & 19.09 & 18.89 & 19.01 & 18.55 & 18.61 & 19.03 & 18.58 & 18.60 & 21.01 & 19.68 & 19.03 & 18.65 & 18.3\\
  132  & 205.7935 & 26.24758  & 0.084 & 2 & 19.88 & 19.41 & 19.26 & 18.89 & 19.01 & 18.52 & 18.26 & 18.22 & 17.98 & 17.78 & 17.86 & 17.78 & 17.75 & 17.69 & 17.76 & 20.38 & 18.75 & 18.16 & 17.88 & 17.7\\
  133  & 205.8046 & 26.45557  & 0.064 & 1 & 21.65 & 20.76 & 20.02 & 19.54 & 19.14 & 19.22 & 19.22 & 18.82 & 18.71 & 18.64 & 18.59 & 18.41 & 18.49 & 18.25 & 18.04 & 20.95 & 19.51 & 18.85 & 18.54 & 18.4\\
  134  & 205.8136 & 26.16733  & 0.068 & 1 & 18.68 & 17.80 & 17.30 & 16.68 & 16.22 & 16.08 & 15.75 & 15.59 & 15.39 & 15.29 & 15.19 & 15.02 & 14.99 & 14.92 & 14.76 & 18.26 & 16.45 & 15.58 & 15.17 & 14.8\\
  135  & 205.8152 & 26.24117  & 0.079 & 3 & 20.37 & 19.78 & 19.63 & 19.31 & 18.85 & 18.79 & 18.48 & 18.39 & 18.17 & 18.32 & 18.16 & 17.88 & 17.83 & 17.97 & 17.84 & 20.56 & 19.15 & 18.45 & 18.09 & 17.9\\
  136  & 205.8347 & 26.32631  & 0.067 & 3 & 18.86 & 18.25 & 17.82 & 17.44 & 17.19 & 17.05 & 16.74 & 16.65 & 16.45 & 16.31 & 16.31 & 16.21 & 16.17 & 16.11 & 15.89 & 18.65 & 17.31 & 16.64 & 16.28 & 16.0\\
  137  & 205.8673 & 26.80032  & 0.068 & 4 & 20.72 & 20.90 & 20.22 & 19.91 & 19.44 & 19.44 & 19.23 & 19.32 & 99.00 & 18.83 & 19.01 & 18.68 & 99.00 & 18.66 & 18.99 & 20.86 & 19.88 & 19.27 & 18.98 & 18.6\\
  138  & 205.8712 & 25.97598  & 0.084 & 4 & 19.80 & 19.26 & 19.04 & 18.72 & 18.25 & 99.00 & 18.02 & 17.91 & 17.67 & 17.41 & 17.49 & 17.37 & 17.35 & 99.00 & 99.00 & 19.67 & 18.45 & 17.84 & 17.51 & 17.3\\
  139  & 205.8717 & 26.68881  & 0.079 & 1 & 21.39 & 20.82 & 20.54 & 20.37 & 19.63 & 19.46 & 19.25 & 19.19 & 18.83 & 19.83 & 18.72 & 18.82 & 18.35 & 18.50 & 18.28 & 21.86 & 19.96 & 19.17 & 18.83 & 18.5\\
  140  & 205.8754 & 26.48968  & 0.087 & 6 & 22.33 & 20.44 & 20.28 & 19.80 & 19.46 & 19.24 & 19.40 & 19.26 & 18.90 & 18.58 & 18.93 & 18.59 & 18.66 & 18.90 & 18.90 & 20.42 & 19.44 & 19.20 & 18.88 & 18.5\\
  141  & 205.8754 & 26.62725  & 0.076 & 2 & 21.11 & 20.24 & 19.77 & 19.36 & 18.80 & 18.57 & 18.22 & 18.16 & 17.84 & 17.78 & 17.58 & 17.43 & 17.42 & 17.34 & 17.07 & 21.17 & 19.10 & 18.09 & 17.63 & 17.2\\
  142  & 205.8787 & 26.78840  & 0.060 & 3 & 21.50 & 20.36 & 20.07 & 20.65 & 20.33 & 19.76 & 20.43 & 19.35 & 99.00 & 19.02 & 19.30 & 18.94 & 99.00 & 19.11 & 18.77 & 21.48 & 20.16 & 19.40 & 19.18 & 18.8\\
  143  & 205.8846 & 26.56603  & 0.071 & 4 & 19.70 & 19.35 & 18.71 & 18.56 & 18.26 & 18.15 & 17.97 & 17.86 & 17.60 & 17.56 & 17.46 & 17.35 & 17.29 & 17.23 & 17.18 & 19.52 & 18.42 & 17.85 & 17.45 & 17.2\\
  144  & 205.9063 & 26.61990  & 0.082 & 3 & 20.38 & 19.71 & 19.20 & 18.89 & 18.41 & 18.23 & 17.95 & 17.81 & 17.61 & 17.50 & 17.37 & 17.32 & 17.36 & 17.16 & 17.03 & 20.13 & 18.63 & 17.81 & 17.36 & 17.0\\
  145  & 205.9949 & 26.41492  & 0.079 & 1 & 21.25 & 20.41 & 19.39 & 18.44 & 18.26 & 18.11 & 17.52 & 17.31 & 17.08 & 17.01 & 16.86 & 16.78 & 16.72 & 16.64 & 16.61 & 19.16 & 15.95 & 18.28 & 15.87 & 14.5\\
  146  & 205.9979 & 26.20461  & 0.071 & 4 & 19.99 & 19.82 & 19.13 & 18.83 & 18.47 & 18.46 & 18.11 & 18.09 & 17.91 & 99.00 & 17.75 & 99.00 & 99.00 & 17.76 & 17.62 & 20.03 & 18.75 & 18.08 & 17.75 & 17.5\\
  \noalign{\smallskip}\hline
  \end{tabular}
\end{table}

\begin{table}[]
\caption{Catalog of spectroscopically confirmed member galaxies in A1775}
\label{table3}
\vspace{3mm}
\def\baselinestretch{1.0}
\centering \tiny
\tabcolsep 1.5mm
\begin{tabular}{ccccl|crcccl}   \hline
\noalign{\smallskip}
 No. & R.A. & Decl. & $z_{sp}$ &
Comments & &No. & R.A.& Decl. & $z_{sp}$ & Comments \\
\noalign{\smallskip}   \hline \noalign{\smallskip}
  1 & 13  39  52.04 & 26  10  13.3 & 0.0727 & a, B  & &   77 & 13  42  02.46 & 26  20  43.3 & 0.0750 & a, B  \\
  2 & 13  39  54.78 & 26  23  12.9 & 0.0747 & a, B  & &   78 & 13  42  02.84 & 26  21  38.3 & 0.0753 & a, B  \\
  3 & 13  39  59.18 & 26  09  02.0 & 0.0740 & a, B  & &   79 & 13  42  05.03 & 26  36  05.6 & 0.0764 & a, B  \\
  4 & 13  40  08.69 & 26  10  38.2 & 0.0652 & a, A  & &   80 & 13  42  05.13 & 26  34  49.3 & 0.0756 & a, B  \\
  5 & 13  40  11.08 & 26  03  30.9 & 0.0778 & a, B  & &   81 & 13  42  06.02 & 26  20  10.3 & 0.0771 & a, B  \\
  6 & 13  40  12.19 & 26  14  52.4 & 0.0764 & a, B  & &   82 & 13  42  06.69 & 26  16  19.3 & 0.0760 & a, B  \\
  7 & 13  40  13.57 & 26  33  54.7 & 0.0831 & c, B  & &   83 & 13  42  06.83 & 26  22  14.3 & 0.0771 & a, B  \\
  8 & 13  40  13.90 & 26  33  52.1 & 0.0758 & a, B  & &   84 & 13  42  09.78 & 26  33  44.8 & 0.0741 & a, B  \\
  9 & 13  40  20.19 & 26  09  39.7 & 0.0729 & a, B  & &   85 & 13  42  10.56 & 26  09  12.6 & 0.0661 & a, A  \\
 10 & 13  40  30.67 & 26  39  39.7 & 0.0769 & a, B  & &   86 & 13  42  10.86 & 26  20  06.4 & 0.0779 & a, B  \\
 11 & 13  40  31.05 & 26  18  26.6 & 0.0625 & c, A  & &   87 & 13  42  14.45 & 26  13  20.0 & 0.0734 & a, B  \\
 12 & 13  40  35.35 & 26  00  53.6 & 0.0755 & a, B  & &   88 & 13  42  16.88 & 26  33  33.4 & 0.0765 & a, B  \\
 13 & 13  40  38.40 & 26  20  11.0 & 0.0758 & a, B  & &   89 & 13  42  18.28 & 26  19  20.3 & 0.0753 & a, B  \\
 14 & 13  40  41.21 & 26  36  55.4 & 0.0649 & a, A  & &   90 & 13  42  20.84 & 26  11  50.5 & 0.0661 & a, A  \\
 15 & 13  40  42.35 & 26  24  38.2 & 0.0750 & a, B  & &   91 & 13  42  24.62 & 26  28  23.5 & 0.0659 & a, A  \\
 16 & 13  40  44.21 & 26  13  44.7 & 0.0775 & a, B  & &   92 & 13  42  25.61 & 26  12  44.8 & 0.0658 & a, A  \\
 17 & 13  40  44.89 & 26  11  11.2 & 0.0645 & a, A  & &   93 & 13  42  25.62 & 26  26  27.3 & 0.0716 & a, B  \\
 18 & 13  40  46.85 & 26  31  41.7 & 0.0690 & a, A  & &   94 & 13  42  26.23 & 26  32  12.6 & 0.0763 & a, B  \\
 19 & 13  40  47.89 & 26  11  43.1 & 0.0756 & c, B  & &   95 & 13  42  29.82 & 25  58  19.6 & 0.0735 & a, B  \\
 20 & 13  40  48.92 & 26  04  17.1 & 0.0656 & b, A  & &   96 & 13  42  31.87 & 26  29  14.0 & 0.0655 & a, A  \\
 21 & 13  40  49.16 & 26  29  15.5 & 0.0648 & a, A  & &   97 & 13  42  33.38 & 26  37  31.8 & 0.0759 & a, B  \\
 22 & 13  40  49.74 & 26  03  51.5 & 0.0650 & d, A  & &   98 & 13  42  35.66 & 26  15  34.0 & 0.0639 & a, A  \\
 23 & 13  40  50.52 & 26  05  54.6 & 0.0652 & a, A  & &   99 & 13  42  35.87 & 26  23  02.1 & 0.0771 & a, B  \\
 24 & 13  40  53.43 & 26  21  00.8 & 0.0775 & a, B  & &  100 & 13  42  37.12 & 26  34  38.3 & 0.0731 & a, B  \\
 25 & 13  40  55.55 & 26  17  57.0 & 0.0739 & a, B  & &  101 & 13  42  39.43 & 26  25  54.1 & 0.0736 & a, B  \\
 26 & 13  40  55.60 & 26  15  18.3 & 0.0752 & a, B  & &  102 & 13  42  39.87 & 26  09  35.7 & 0.0654 & a, A  \\
 27 & 13  40  55.95 & 26  24  51.8 & 0.0759 & a, B  & &  103 & 13  42  41.39 & 26  14  23.3 & 0.0664 & a, A  \\
 28 & 13  40  56.59 & 26  29  12.3 & 0.0750 & a, B  & &  104 & 13  42  41.46 & 26  28  13.3 & 0.0645 & a, A  \\
 29 & 13  40  57.07 & 26  10  21.6 & 0.0634 & a, A  & &  105 & 13  42  41.99 & 26  14  23.2 & 0.0654 & b, A  \\
 30 & 13  40  58.86 & 26  35  05.2 & 0.0759 & a, B  & &  106 & 13  42  42.02 & 26  17  09.7 & 0.0677 & a, A  \\
 31 & 13  41  00.52 & 26  29  07.2 & 0.0760 & b, B  & &  107 & 13  42  43.22 & 26  12  16.0 & 0.0658 & a, A  \\
 32 & 13  41  06.32 & 26  36  52.6 & 0.0773 & a, B  & &  108 & 13  42  44.86 & 26  10  58.1 & 0.0663 & a, A  \\
 33 & 13  41  06.38 & 26  15  31.5 & 0.0805 & a, B  & &  109 & 13  42  46.36 & 26  17  30.9 & 0.0665 & a, A  \\
 34 & 13  41  07.95 & 26  30  48.3 & 0.0703 & a, B  & &  110 & 13  42  47.07 & 26  18  33.6 & 0.0739 & a, B  \\
 35 & 13  41  13.43 & 26  29  33.2 & 0.0734 & a, B  & &  111 & 13  42  47.17 & 26  21  53.9 & 0.0644 & a, A  \\
 36 & 13  41  16.60 & 26  23  35.8 & 0.0736 & a, B  & &  112 & 13  42  48.34 & 26  45  00.1 & 0.0764 & a, B  \\
 37 & 13  41  17.08 & 26  16  19.4 & 0.0639 & a, A  & &  113 & 13  42  50.81 & 26  16  40.5 & 0.0731 & a, B  \\
 38 & 13  41  18.04 & 26  47  51.4 & 0.0765 & a, B  & &  114 & 13  42  52.17 & 26  12  10.7 & 0.0643 & a, A  \\
 39 & 13  41  19.72 & 26  21  14.8 & 0.0734 & a, B  & &  115 & 13  42  52.86 & 26  35  33.3 & 0.0696 & a, B  \\
 40 & 13  41  20.05 & 25  53  25.7 & 0.0735 & a, B  & &  116 & 13  42  53.79 & 26  08  37.8 & 0.0719 & a, B  \\
 41 & 13  41  20.11 & 26  30  06.7 & 0.0707 & a, B  & &  117 & 13  42  54.30 & 26  13  57.1 & 0.0743 & a, B  \\
 42 & 13  41  22.94 & 26  28  09.4 & 0.0795 & a, B  & &  118 & 13  42  55.45 & 26  16  28.0 & 0.0763 & a, B  \\
 43 & 13  41  25.93 & 26  33  12.4 & 0.0747 & a, B  & &  119 & 13  42  57.45 & 26  25  30.1 & 0.0733 & a, B  \\
 44 & 13  41  27.21 & 26  25  41.8 & 0.0749 & a, B  & &  120 & 13  42  57.86 & 26  02  25.0 & 0.0659 & a, A  \\
 45 & 13  41  29.17 & 26  10  08.2 & 0.0768 & a, B  & &  121 & 13  42  57.99 & 25  56  23.2 & 0.0632 & a, A  \\
 46 & 13  41  32.25 & 26  01  27.6 & 0.0662 & a, A  & &  122 & 13  42  59.01 & 26  15  49.3 & 0.0673 & a, A  \\
 47 & 13  41  35.18 & 26  21  24.0 & 0.0726 & a, B  & &  123 & 13  43  03.34 & 26  45  08.5 & 0.0757 & a, B  \\
 48 & 13  41  36.69 & 26  41  26.4 & 0.0779 & a, B  & &  124 & 13  43  07.49 & 26  03  26.9 & 0.0672 & a, A  \\
 49 & 13  41  37.42 & 26  20  27.6 & 0.0781 & a, B  & &  125 & 13  43  09.64 & 26  10  40.6 & 0.0644 & a, A  \\
 50 & 13  41  38.59 & 26  06  09.7 & 0.0640 & a, A  & &  126 & 13  43  10.24 & 25  56 59.1  & 0.0612 & a, A  \\
 51 & 13  41  38.90 & 26  28  47.4 & 0.0769 & a, B  & &  127 & 13  43  12.10 & 26  02 24.5  & 0.0616 & a, A  \\
 52 & 13  41  38.95 & 26  14  27.9 & 0.0707 & a, B  & &  128 & 13  43  14.77 & 26  00  57.4 & 0.0846 & a, B  \\
 53 & 13  41  39.52 & 26  09  22.4 & 0.0749 & a, B  & &  129 & 13  43  15.32 & 26  39  03.9 & 0.0760 & a, B  \\
 54 & 13  41  39.97 & 26  25  11.4 & 0.0648 & a, A  & &  130 & 13  43  15.55 & 26  09  57.8 & 0.0647 & b, A  \\
 55 & 13  41  40.11 & 26  29  40.3 & 0.0753 & a, B  & &  131 & 13  43  15.84 & 26  09  52.8 & 0.0642 & a, A  \\
 56 & 13  41  41.24 & 26  17  45.2 & 0.0724 & a, B  & &  132 & 13  43  17.21 & 26  30  34.8 & 0.0750 & a, B  \\
 57 & 13  41  42.42 & 26  15  31.9 & 0.0738 & a, B  & &  133 & 13  43  17.28 & 26  19  43.2 & 0.0775 & a, B  \\
 58 & 13  41  43.81 & 26  17  36.2 & 0.0757 & b, B  & &  134 & 13  43  18.32 & 26  14  07.1 & 0.0660 & a, A  \\
 59 & 13  41  46.49 & 26  19  35.0 & 0.0744 & a, B  & &  135 & 13  43  22.85 & 26  26  36.9 & 0.0769 & a, B  \\
 60 & 13  41  47.14 & 26  27  34.3 & 0.0779 & a, B  & &  136 & 13  43  25.33 & 26  13  10.6 & 0.0742 & a, B  \\
 61 & 13  41  47.19 & 26  22  51.4 & 0.0772 & a, B  & &  137 & 13  43  29.13 & 26  29  36.7 & 0.0764 & a, B  \\
 62 & 13  41  49.14 & 26  22  24.5 & 0.0757 & f, B  & &  138 & 13  43  29.44 & 26  09  32.8 & 0.0661 & a, A  \\
 63 & 13  41  49.83 & 26  08  41.6 & 0.0732 & a, B  & &  139 & 13  43  30.99 & 25  55  52.9 & 0.0642 & a, A  \\
 64 & 13  41  50.45 & 26  22  13.0 & 0.0694 & a, A  & &  140 & 13  43  31.08 & 26  42  03.1 & 0.0787 & a, B  \\
 65 & 13  41  50.60 & 26  21  10.6 & 0.0759 & a, B  & &  141 & 13  43  42.26 & 26  15  47.1 & 0.0745 & a, B  \\
 66 & 13  41  50.91 & 26  22  28.3 & 0.0750 & e, B  & &  142 & 13  43  45.77 & 26  20  15.8 & 0.0735 & a, B  \\
 67 & 13  41  51.81 & 26  05  56.9 & 0.0661 & a, A  & &  143 & 13  43  49.09 & 26  09  27.0 & 0.0833 & a, B  \\
 68 & 13  41  53.39 & 26  25  12.6 & 0.0773 & a, B  & &  144 & 13  44  00.22 & 26  44  42.8 & 0.0712 & a, B  \\
 69 & 13  41  54.08 & 26  22  50.8 & 0.0758 & a, B  & &  145 & 13  44  01.90 & 25  56  28.3 & 0.0620 & a, A  \\
 70 & 13  41  55.13 & 26  20  35.7 & 0.0736 & a, B  & &  146 & 13  44  02.81 & 26  09  05.7 & 0.0666 & a, A  \\
 71 & 13  41  56.48 & 26  27  16.4 & 0.0736 & a, B  & &  147 & 13  44  02.86 & 26  06  35.8 & 0.0672 & a, A  \\
 72 & 13  41  57.71 & 26  24  17.1 & 0.0767 & a, B  & &  148 & 13  44  03.33 & 26  18  12.8 & 0.0653 & a, A  \\
 73 & 13  41  58.54 & 26  21  46.3 & 0.0748 & a, B  & &  149 & 13  44  05.41 & 26  20  17.7 & 0.0770 & a, B  \\
 74 & 13  41  59.50 & 26  23  18.3 & 0.0734 & a, B  & &  150 & 13  44  08.67 & 26  20  56.1 & 0.0651 & a, A  \\
 75 & 13  42  00.48 & 26  18  38.2 & 0.0782 & a, B  & &  151 & 13  44  09.92 & 26  48  02.9 & 0.0734 & a, B  \\
 76 & 13  42  02.30 & 26  10  42.8 & 0.0744 & a, B  & &   &  &   &  &  \\
\noalign{\smallskip}\hline
\end{tabular}
\begin{minipage}[]{125mm}
\vspace {3mm} \footnotesize References:(a)Extract from SDSS; (b)
Oegerle Hill, and Fitchett (1995);
 (c) Kirshner et al. (1983); (d) Crampton et al. (1992); (e) Beckmann et al. (2003);
 (f)Davoust \& Considere (1995). (A) Members of A1775A; (B) Members of A1775B.
\end{minipage}
\end{table}

\begin{table}[]
\caption{Result of~$\kappa$-test~for 151 spectroscopically
confirmed member galaxies}
\label{table4}
\scriptsize
\begin{center}
\begin{tabular}{ccccccccc}   \hline
\noalign{\smallskip}
Neighbor size $n$ & 5& 6& 7& 8& 9 &10&11 \\
\hline $P(\kappa_n>\kappa_n^{\rm obs})$
&0.3\%& 0.2\%& 1.8\%& 0.5\%& $<0.05\%$ & 0.1\% & 0.2\%\\
\noalign{\smallskip} \hline
\end{tabular}
\end{center}
\end{table}

\end{document}